\documentclass
[preprint,byrevtex,nofootinbib,balancelastpage,titlepage,bibnotes,pre,twocolumn,10pt]{revtex4}%
\usepackage{amsfonts}
\usepackage{amsmath}
\usepackage{amssymb}
\usepackage{graphicx}%
\providecommand{\U}[1]{\protect\rule{.1in}{.1in}}
\ifx\pdfoutput\relax\let\pdfoutput=\undefined\fi
\newcount\msipdfoutput
\ifx\pdfoutput\undefined\else
\ifcase\pdfoutput\else
\ifx\paperwidth\undefined\else
\ifdim\paperheight=0pt\relax\else\pdfpageheight\paperheight\fi
\ifdim\paperwidth=0pt\relax\else\pdfpagewidth\paperwidth\fi
\fi\fi\fi
\begin{document}
\preprint{UATP/1803}
\title{The Role of the Communal Entropy and Free Volume for the Viscosity Divergence
near the Glass Transition:\ A New Conceptual Approach}
\author{P.D. Gujrati}
\email{pdg@uakron.edu}
\affiliation{Department of Physics, Department of Polymer Science, The University of Akron,
Akron, OH 44325}

\begin{abstract}
The conventional approach to study glasses either requires considering the
rapid drop in the excess entropy $\Delta S_{\text{ex}}$ or the free volume
$V_{\text{f}}$. As the two quantities are not directly related to each other,
the viscosity in the two approaches does not diverge at the same temperature,
which casts doubt on the physical \emph{significance of the divergence} and of
the \emph{ideal glass transition }(IG). By invoking a recently developed
nonequilibrium thermodynamics, we identify the instantaneous temperature,
pressure, entropy, etc. and discover the way they relax. We show that by
replacing $\Delta S_{\text{ex}}$ by a properly defined communal entropy
$S^{\text{comm}}$ (not to be confused with the configurational entropy) and
free volume $V_{\text{f}}$, both quantities vanish simultaneously at IG, where
the glass is\emph{ jammed} with no free volume and communal entropy. By
exploiting the fact that there are no thermodynamic singularities in the
entropy of the supercooled liquid at IG, we show that various currently
existing phenomenologies become \emph{unified}.

\end{abstract}
\date{\today}
\maketitle

\section{Introduction\label{marker-Sec-Introduction}}

\subsection{ Glass as a Time-dependent Nonequilibrium Macrostate}

\subsubsection{Glass Transition Temperature}

Vitrification \cite{Zallen} is a prime example of an irreversible process
going on at low temperatures or high pressures. It is commonly believed now
that almost all materials including organic and inorganic substances, man-made
polymers, metals, plastics, biomaterials, drugs, etc. can be turned into a
glass (or vitrified) by exploiting a suitable technique or techniques. Even
though naturally occurring glasses such as volcanic glasses\footnote{These
glasses formed by lava are mostly of Paleogene age or younger and
spontaneously devitrify to crystallite aggregates in time periods that are
extremely short by geologic standards but extremely long by human standards.
This is consistent with our understanding that glasses are \emph{unstable}
over extremely long periods.} have been known for a long time, our
understanding of them is far from complete, mainly because they exhibit a
\emph{duality} in their properties, some of which appear liquidlike at long
times, while others appear solidlike at short times; see Moynihan, et al in
\cite{Goldstein}. As observed using various spectroscopic techniques,
molecular motions in liquids became progressively slower as vitrification is
approached, with a characteristic time increasing from nanoseconds to beyond
feasible experimental time scales time $\tau_{\text{exp}}$ (the inverse of the
probed frequency $\omega$) of the order of $10^{2}$ s or $10^{5}$ s$\simeq$
one day. This results in an operational definition of a range of temperatures
$T_{0\text{g}}$-to-$T_{0\text{G}}$ for the glass transition; however,
following convention, we simply use $T_{0\text{g}}$ to denote this range. It
is found to depend on the rate of approach to the transition. The slow glassy
dynamics is thought to occur because particles form cooperative groups of
increasing sizes, which then identify a \emph{correlated length} as the ratio
$\zeta_{0}\equiv P_{0}/T_{0}$ of pressure to temperature increases. This is
the idea behind the celebrated Adam-Gibbs theory of glass transition; see
later. Similar time dependence is also found in a class of disordered magnets,
commonly know as spin glass. It is found that a spin glass exhibits an
exponentially large number of metastable states below the (spin-)glass
transition temperature, which are specifically determined by the presence of
\emph{frustration}, where by frustration is meant the system's inability to
simultaneously minimize the (sometimes competing) interaction energy between
its constituents. The similarity between a glass and a spin glass has
suggested frustration to be a deciding factor for a regular glass, although
the situation is not very clear.

\subsubsection{Fast and Slow Modes}

\begin{figure}
[ptb]
\begin{center}
\includegraphics[
height=2.4076in,
width=2.444in
]%
{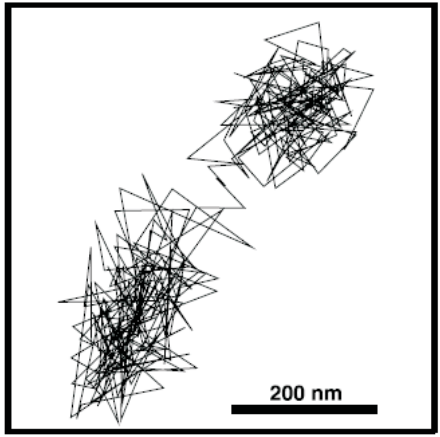}%
\caption{The 2-d projection of a typical 3-d trajectory for 100 min for
$\phi=0.56$. Most of the time, particles are confined to their cages.
Occasionally, a particle will move a long distance and get trapped in another
cage. In the figure, the particle took $500$ s to shift its position.
Reprinted with permission from E.R. Weeks, et al, Science \textbf{287}, 627
(2000).}%
\label{Fig_Simulation}%
\end{center}
\end{figure}

The conclusion from several studies on glasses and spin glasses is that there
are at least \emph{two distinct} (widely separated) time scales governing
\emph{fast} and \emph{slow} modes in glasses. The fast modes refer to the
localized oscillations of the particles (atoms, molecules, segments, monomers,
etc.) in the cells or cages formed by neighboring particles, and the slow
modes refer to the translation and diffusion (translation over long distances
\cite{Zallen} of particles. In Fig. \ref{Fig_Simulation}, we show a
two-dimensional projection of a three-dimension trajectory of a colloidal
particle at high enough density, where the particle oscillates within its cage
for a long time before leaving to a new cage in which it oscillates again to
leave it later on, and so on. These two distinct modes distinguish the glass
from other nonequilibrium states where there is only one single mode.
Consequently, the study of vitrification continues to attract researchers to
this date because of its incomplete and some times controversial
understanding.%
%TCIMACRO{\FRAME{ftbpFU}{2.444in}{2.4076in}{0pt}{\Qcb{The 2-d projection of a
%typical 3-d trajectory for 100 min for $\phi=0.56$. Most of the time,
%particles are confined to their cages. Occasionally, a particle will move a
%long distance and get trapped in another cage. In the figure, the particle
%took $500$ s to shift its position. Reprinted with permission from E.R. Weeks,
%et al, Science \QTR{bf}{287}, 627 (2000).}}{\Qlb{Fig_Simulation}%
%}{Weitz-Simulation.eps}{\special{ language "Scientific Word";
%type "GRAPHIC";  maintain-aspect-ratio TRUE;  display "USEDEF";
%valid_file "F";  width 2.444in;  height 2.4076in;  depth 0pt;
%original-width 4.0248in;  original-height 3.9669in;  cropleft "0";
%croptop "1";  cropright "1";  cropbottom "0";
%filename 'Weitz-Simulation.eps';file-properties "NPEU";}}}%
%BeginExpansion
%EndExpansion
\qquad

\subsubsection{What is a Glass?}

As a physicist, the first thing we need to ask is:\ What is glass? The simple
and most common answer is that glass or glassy state (GS) is a
\emph{time-dependent} \emph{nonequilibrium} state of matter
\cite{Kauzmann,Goldstein} found at low temperatures $T_{0}$ and/or high
pressures $P_{0}$ of the medium that undergoes glass transition (GT)\ from a
relatively brittle solid (glass) into a molten state as the $\zeta_{0}$
decreases. The striking feature of a glass which distinguishes it from an
equilibrium crystal is that the former has much higher potential energies than
the latter \cite{Landau}. Even at absolute zero, there remains a non-zero gap
in the energies of the crystal and the glass. Thus, a glass can be thought of
as a macrostate which has trapped a lot of frozen defects that were present in
the liquid at the melting point after the crystal melts \cite{Gujrati-Book}.

A glass is most commonly obtained in the laboratory by cooling a
time-dependent supercooled liquid (SCL) to a temperature lower than the
temperature at which the latter has fallen out of equilibrium. The genuine
equilibrium state, which by definition is also a \emph{stationary state},
corresponds to a crystalline state (CR). This does not mean that SCL can never
be treated as an equilibrium state; its stationary limit, to be called the
equilibrium SCL (ESCL), can be treated as an equilibrium state under the
constraint that only \emph{disordered} microstates are considered
\cite{Gujrati-Book}. We denote the ensemble or the space of disordered
microstates by $\mathcal{D}$, the \emph{disordered state space}, and refer to
a constraint formalism as a \emph{restricted ensemble} in this work to
distinguish it from an unrestricted ensemble in which all microstates are
considered. As expected, the corresponding entropy $S_{\text{ESCL}}$ has its
maximum value for given extensive arguments (the energy $E$, volume $V$,
number of particles $N$, etc.). To make a clear distinction, all possible
time-dependent SCL states will be called nonequilibrium SCL (NESCL) states. It
can be shown that the entropy $S(t)$ of any system in any arbitrary state
(equilibrium or not) is given by \cite{Gujrati-Entropy}%
\begin{equation}
S(t)=-%
%TCIMACRO{\tsum }%
%BeginExpansion
{\textstyle\sum}
%EndExpansion
p_{i}\ln p_{i}\geq0 \label{General-Entropy}%
\end{equation}
in terms of microstate probabilities $p_{i}$ of the \emph{allowed} $i$th
microstate \emph{provided} $S\left(  t\right)  $ is an \emph{extensive}
quantity; no other assumption is required for the derivation. This is the
standard formulation of entropy that is applicable to both ensembles and to
any nonequilibrium state. These entropies achieve their unique maximum value
$S_{\text{ESCL}}$ ($S_{\text{CR}}$) in the stationary limit in the restricted
(unrestricted) ensemble. In the following, we will use SCL to refer to both
nonequilibrium and equilibrium SCLs.

When the internal energy consists of two \emph{independent} contributions due
to positions (configurations) and momenta (kinetic energy) of the particles,
respectively, then the microstate probability becomes a product of
probabilities due to the independent contributions. It then follows trivially
from Eq. (\ref{General-Entropy}) that the entropy of the system can be divided
into two independent terms: the configurational entropy $S^{\text{conf}}(t)$
due to the positions and interactions (which we take to be independent of
momenta) and the entropy $S^{\text{kin}}(t)$ due to the kinetic energy,
respectively%
\begin{equation}
S(t)=S^{\text{conf}}(t)+S^{\text{kin}}(t). \label{CK-Partition}%
\end{equation}
Such a separation is possible for any macrostate of the system
\cite{Gujrati-Book}. Each independent entropy contribution above must satisfy
the second law according to which the entropy continues to increase until
equilibrium is reached. However, the configurational entropy in the glass
community sometimes refers to the communal entropy. It is the entropy due to
the \emph{deconfinement} of the system from their cells. We have discussed
this issue in Secs. 10.1.5.2\ and 10.4.1 in Ref. \cite{Gujrati-Book}. Here, we
will focus on the configurational entropy $S^{\text{conf}}(t)$, see below, for
which we will assume the second law to hold. It is given by Eq.
(\ref{General-Entropy}) except that $p_{i}$ now refers to a microstate in the
configuration space. Therefore, it must also be non-negative. A related
quantity of interest is the measured excess entropy%
\begin{equation}
S_{\text{ex}}(T_{0},t)\equiv S_{\text{SCL}}(T_{0},t)-S_{\text{CR}}%
(T_{0}),T_{0}<T_{0\text{m}}; \label{Ex_Partition}%
\end{equation}
here, $T_{0}$ denotes the temperature of the surroundings
(medium)\footnote{From now on, we will use "medium" instead of
"surroundings."} and $T_{0\text{m}}$ is the melting temperature.

The layout of the paper is as follows. In the next section, we briefly review
some important concepts and topics used in the paper, which is followed by a
brief review of glassy phenomenology. Nonequilibrium thermodynamics is
discussed in Sect. III, and its consequences for glasses is considered in
Sect. IV. The topics of free volume and the communal entropy is taken up in
Sect. V. We review some of the established theories in Sect. VI. A simple
model of a nonequilibrium temperature is discussed in Sect. VII and is applied
in Sect. VIII to give a thermodynamic justification of the Tool-Narayanaswamy
equation. The final section contains a discussion of the results and the
limitation of the approach.

\section{Brief Review of Important Concepts and Topics}

\subsection{Entropy Crisis and the Ideal Glass Transition}

The excess entropy $S_{\text{ex}}(T_{0},t)$ has played a very pivotal role in
the field of glass transition. It was found to exhibit a rapid drop below the
melting temperature $T_{0\text{m}}$ by Kauzmann \cite{Kauzmann} for many
systems; a smooth extrapolation to lower temperatures shows that it eventually
vanishes at some temperature $T=$ $T_{0\text{K}}<T_{0\text{m}}.$ It will
become \emph{negative if extrapolated} to lower temperatures.\footnote{It
should be stressed that there is no thermodynamic requirement for
$S_{\text{ex}}(T_{0})$ to be non-negative. There are physical systems like
He$^{4}$ in which $S_{\text{ex}}(T_{0})$ can become negative at low
temperatures. This means that the vanishing of $S_{\text{ex}}(T_{0}%
)$\ \emph{cannot} be an argument for a glass transition but the vanishing of
$S_{\text{ESCL}}^{\text{comm}}(T_{0})$\ willl be.},\footnote{One can consider
many different kinds of entropy, all of which will vanish, most probably at
different $\zeta_{0}$, under extrapolation. Similarly, various definitions of
the free volume will also show that they vanish at different $\zeta_{0}$ under
extrapolation. Thus, unless care is taken, the vanishing of entropy and free
volume will appear to be unrelated. This has created a lot of confusion in the
field. However, by carefully defining the free volume and the relevant
(communal) entropy, we can show that their vanishing is simultaneous, and that
they refere to the same unique IG macrostate.} A (kinetic) transition known as
the glass transition (GT) is invoked at a higher temperature $T_{0\text{g}%
}>T_{0\text{K}}$ to avoid this \emph{entropy crisis}, known commonly as the
\emph{Kauzmann paradox} at $T_{0\text{K}}$. In the limit of zero cooling rate
(not accessible in experiments or simulations, but accessible in a theoretical
setup) in the metastable region where SCLs occur, the metastable states will
become stationary, which we have identified above as ESCL. In this limit,, the
glass transition in ESCL at $T_{0\text{K}}$ is known as the \emph{ideal glass
transition} (IGT) to an ideal glass (IG) below $T_{0\text{K}}$.

It should be stressed that there is no thermodynamic requirement for
$S_{\text{ex}}(T)$ to be non-negative. There are physical systems like
He$^{4}$ in which $S_{\text{ex}}(T)$ can become negative at low temperatures;
see also Ref. \cite{Gujrati-Book}. On the other hand, the SCL\ configurational
entropy $S_{\text{SCL}}^{\text{conf}}(T)$ can never be negative. This means
that $S_{\text{SCL}}^{\text{conf}}(T)$ and $S_{\text{ex}}(T)$\ are \emph{not}
close under all relevant conditions. If there is any hope of finding a
thermodynamic basis for the glass transition occurring in SCL, we must look
for the violation of the condition $S_{\text{SCL}}^{\text{conf}}(T)$ $\geq0$
and not of $S_{\text{ex}}(T)$ $\geq0$. Thus, in the following, we will only
consider the configurational entropy $S_{\text{SCL}}^{\text{conf}}(T)$. We
will interpret the entropy crisis in this work to signify the reality
condition violation $S_{\text{SCL}}^{\text{conf}}(T)$ $\geq0$, and denote the
temperature by $T_{0\text{K}}$, the Kauzmann temperature, where the violation
begins to occur as the temperature is reduced. A similar Kauzmann pressure
$P_{0\text{K}}$ can be identified where the reality condition is violated as
the pressure is increased. We will collectively call them Kauzmann points.
However, we will mostly consider the Kauzmann temperature in this work.

In the rest of the paper, we will simplify the notation \ and no longer
exhibit the superscript configurational in the entropy. Thus, $S(T_{0},t)$
from on will refer to $S^{\text{conf}}(T_{0},t)$. As the kinetic energy is no
longer going to be considered, the energy $E$ will now represent the potential energy.

Normally, the time dependence in experimentally measured $S_{\text{ex}}%
(T_{0},t)$ and $S_{\text{SCL}}(T_{0},t)$ is not taken into account if the
interest is to explain the rapid entropy variation with $T_{0}$ or $P_{0}$.
With the presence of $t$, one can hope to also account for the possibility of
relaxation in the time domain such as for aging. However, the glassy dynamics
in the time domain is too complex as it contains a variety of different
processes ($\alpha$- and $\beta$- relaxations, Johari-Goldstein relaxation and
its connection with the $\beta$- relaxation, dependence of relaxation on the
waiting time, memory effect, etc.) and their origin and physical significance
are still debated. As these processes are found in a variety of systems, they
are believed to be generic to all glass formers and the hope is that there
must be some universal explanation of these processes that are independent of
the particular properties of the glass considered. Our goal in this paper is
to focus on the thermodynamics of glasses. Therefore, we will not be
interested in the actual form of the temporal variation except to note that
the most common empirical law, valid in a limited domain, is found to be the
Kohlrausch stretched exponential \cite{Ngai}%
\begin{equation}
q(T_{0},t)=q_{0}\exp(-t/\tau_{\text{eff}})^{\beta} \label{Kohlrausch-Form}%
\end{equation}
for some physical observable like the volume, viscosity, refractive index,
elastic constants, etc. under fixed external conditions such as $T_{0},P_{0}$;
here $\tau_{\text{eff}}$ represents some average relaxation time, and $q_{0}$
and $\tau_{\text{eff}}$ must be state dependent quantities.

\subsection{Free Volume}

In this paper, we are only concerned with the motion of individual particles,
which makes the cell theory of liquids very appealing \cite{Zallen}. The
localized oscillatory motion experienced by a particle occurs about the
minimum of the potential $\varphi$ generated by its neighbors; the latter
define the cell. (The set of minima from all the cells determines what is
nowadays called the\emph{ inherent structure} IS). While in a crystal, such a
motion occurs at all times unless there are interstitial vacancies. In a
glass, such a motion occurs at short time and endows the glass with solid-like
properties; see Fig. \ref{Fig_Simulation}. But at long times, there occurs
uninhibited translation and diffusion, superimposed on the oscillatory motion
controlled by $\varphi$ of its continuously changing new neighbors
\cite{Zallen}. This gives the glass a liquid-like property, whose central
feature is the mobility of the particle. At high densities, the neighbors
impede the motion in almost all directions and the motion become confined
within the cell or cage with little or no diffusion. At low densities, the
particle can move almost freely in any direction and diffusion occurs.
However, the presence of chemical boding requires the whole molecule to move
together. We are interested in the dense phase where we have both motions
possible. The ability to move long distances requires, what is vaguely termed
the \emph{free volume} $V_{\text{f}}$, and which is communally shared by many
particles. The potential felt by the particle in its cell endows the particle
with what can be termed the interaction volume $v_{\text{i}}$ per particle. It
is the volume necessary to execute its oscillatory motion in this potential.
In terms of the interaction volume $V_{\text{i}}\equiv Nv_{\text{i}}$, we have%
\begin{equation}
V=V_{\text{i}}+V_{\text{f}}; \label{Volume-Partition}%
\end{equation}
both components are functions of state variables like temperature, pressure,
etc;\footnote{Many workers use the concept of occupied volume $V_{\text{o}%
}\equiv Nv_{\text{o}}$. However, the concept is not without any ambiguity
\cite{Matsuoka}. For some, $v_{\text{o}}$ is just the van der Waals volume,
also known as the molecular volume $v_{\text{m}}$. It is just a parameter of
the liquid. In contrast, the interaction volume is theoretically well defined
as shown later. To inquire if they are the same is meeaningless. As a van der
Waals volume is merely a parameter, $V_{\text{o}}$ is just a constant equal to
$Nv_{\text{m}}$. Such a definition will not account for the temperature
variation of the ideal glass volume in Fig. \ref{Fig_Glass_Transition_Time}.
Our definition allows for such a variation. With our definition, as we will
discuss later, the communal entropy vanishes and the free volume vanish
simultaneously, which is what we expect if the ideal glass is a unique
macrostate of the system.} see also Ref. \cite{Matsuoka} The interaction
volume is determined by the local modes within the cells, whereas the free
volume is determined by the translational motion of the particles inside and
outside the cell that give rise to its \emph{fluidity}. The cell potential
must gradually become very steep as the particle gets closer to the neighbors.
At low temperatures, these steep portions of the potential have almost no
chance to be explored by the particle. Therefore, the interaction volume
$v_{\text{i}}$ must be usually smaller than the cell volume $\Delta$. Their
difference $\Delta-$ $v_{\text{i}}$ gives the particle some elbow room to
allow for translation; it is thus included in $V_{\text{f}}$. The
determination of $V_{\text{i}}$ is somewhat technical and will be discussed later.

\subsection{Doolittle Equation, its Generalization and the Free Volume Theory}

According to the Doolittle equation \cite{Doolittle}, the fluidity $\phi$,
which is basically the inverse of the viscosity $\eta$, is given by%
\begin{equation}
\phi\equiv\eta^{-1}=\phi_{0}\exp(-\gamma v_{\text{m}}/v_{\text{f}}),
\label{Doolittle Eq}%
\end{equation}
where $\gamma$\ is a fitting parameter of order unity and $v_{\text{m}}$ is
the molecular volume. As the free volume decreases so does $\phi$ or
$\eta^{-1}$. The glass transition normally occurs when $\eta$ becomes larger
than about $10^{13}$ poise or the the relaxation time becomes of the order
$\tau_{\text{exp}}$; see the%
\begin{figure}
[ptb]
\begin{center}
\includegraphics[
height=2.5131in,
width=3.4852in
]%
{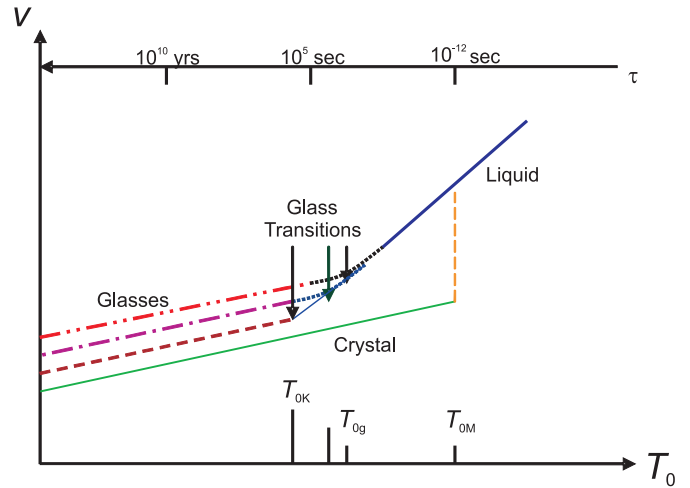}%
\caption{Schematic behavior of the volume as the liquid is cooled. The
freezing transition to the crystal occurs at $T_{0\text{m}}$; the latter
becomes perfectly ordered at absolute zero. If the crystallization is somehow
bypassed, we obtain the supercooled liquid (shown as the continuation of the
liquid), which eventually turns continuously without any discontinuity in the
slope into different glasses at different glass transition temperatures
($T_{0\text{g}}$) depending on the rate of cooling $r$. As $r$ becomes
smaller, $T_{0\text{g}}$ decreases (shown by arrows becoming larger), until
finally it converges to its limit $T_{0\text{K}}$ under infinitely slow
cooling rate but now with a discontinuity in the slope. This limit is called
the \emph{ideal glass transition temperature} or the \emph{Kauzmann
temperature}, and the corresponding glass shown by the dashed curve is called
the \emph{ideal glass}. A similar behavior in the slopes of the densities is
also seen when we increase $P_{0}$ at a fixed $T_{0}$; we merely replace
$T_{0}$ in the figure by $1/P_{0}.$Thus, we can replace the horizontal axis by
$1/\zeta_{0}$. }%
\label{Fig_Glass_Transition_Time}%
\end{center}
\end{figure}
%EndExpansion
upper axis in Fig. \ref{Fig_Glass_Transition_Time}, where we have plotted the
volume per particle $V$ as a function of the temperature $T_{0}$ of the medium
but the following discussion applies equally well to any density function like
the entropy, energy, enthalpy etc. per particle and applies to all glass
forming systems. In the Doolittle equation, the parameters $\gamma$ and
$v_{\text{m}}$ are \emph{constant}. But in general, these parameters must be
functions of the state variables. With this dependence, we will refer to the
above equation as the \emph{generalized Doolittle equation}. It is found that
a linear temperature-dependent $v_{\text{f}}=a(T_{0}-T_{0\text{V}})$,
$T_{0\text{V}}\neq0$, $a$ and $T_{0\text{V}}$ constants, is satisfied only
over a narrow range of the temperature $T_{0}$ for most substances; see Ref.
\cite{Cohen-Grest} for more details. At $T_{0\text{V}}$, $V_{\text{f}}$
vanishes so that there cannot be any translational motion. The system becomes
completely \emph{jammed}. The entropy associated with translational motion,
which we call the communal entropy $S^{\text{comm}}$, must also vanish. (We
identify $S^{\text{comm}}$ in Sec. \ref{Sec._communal_entropy}.)

The vanishing of $V_{\text{f}}$ is reflected in the behavior of the volume;
see the point on the lowest curve at temperature $T_{0\text{K}}$ in Fig.
\ref{Fig_Glass_Transition_Time}, where there is a sharp kink and a
discontinuity in the slope.\footnote{We have identified the temperature as
$T_{0\text{K}}$ rather than $T_{0\text{V}}$ with the anticipated result
obtained later that $T_{0\text{K}}$, the Kauzmann temperature, is identified
as the point where the communal entriopy vaishes is the same as $T_{0\text{V}%
}$, where the free volume vanishes. This is, as it should be, because the
ideal glass shown by the dashed line must be a unique state.} This curve is
the result of extrapolation to obtain ESCL\ shown by the solid line (Liquid).
The presence of free volume above this temperature gives rise to different
expansion coefficients on the two sides of this point. The dashed portion
represents the \emph{ideal glass} (IG). The other two curves represent
laboratory glasses which smoothly emerge out of the solid curve representing
ESCL; the glass transition (GT) occurs at a lower temperature than where the
curves leave ESCL. The free volume does not vanish on these curves. If we
believe that IG is a unique state, though we do not know at present how to
specify it, it must also be the state with vanishing communal entropy (but not
vanishing configurational entropy in (\ref{CK-Partition})). This is shown by
the point at temperature $T_{0\text{K}}$ in Fig. \ref{Fig_entropyglass-Ideal}%
(b), where the communal entropy vanishes. Expecting this identification to be
justified later in the paper, we will henceforth use $T_{0\text{K}}$ for
$T_{0\text{V}}$.\ The free volume picture provides a very nice way to think of
the glass transition \cite{Zallen} with percolation of the free volume as an
important ingredient \cite{Cohen-Grest}.

Similar to the Kauzmann temperature $T_{0\text{K}}$ for isobaric vitrification
at fixed $P_{0}$., a Kauzmann pressure $P_{0\text{K}}$ can also be identified
in ESCL where $S_{\text{ESCL}}^{\text{comm}}(P_{0})$ $=0$ as the pressure is
increased at fixed $T_{0}$. We will collectively call them Kauzmann points.

\section{Generic Glass Phenomenology}

\subsection{Absence of a Singularity in Laboratory Glass}

\begin{figure}
[ptb]
\begin{center}
\includegraphics[
trim=-0.061445in 0.000000in 0.000000in 0.000000in,
height=1.9899in,
width=3.589in
]%
{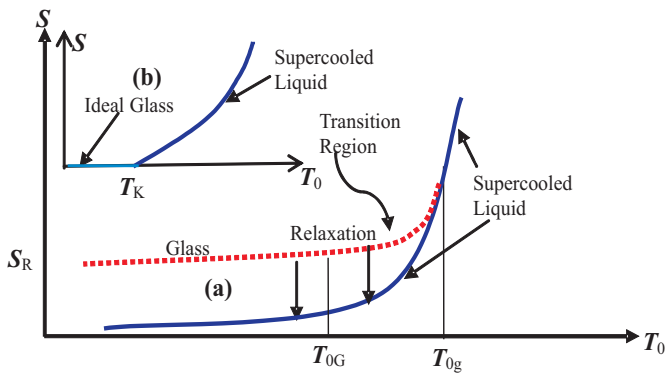}%
\caption{(a) Schematic behavior of the communal entropy of ESCL (solid curve)
and a possible time-dependent supercooled liquid NESCL, which turns into a
glass\ (dotted curve) during vitrification. It is assumed that there is no
ideal glass transition in ESCL. The transition region between $T_{0\text{g}}$
and $T_{0\text{G}}$ has been exaggerated to highlight the point that the glass
transition is not a sharp point.\ For all temperatures $T_{0}<T_{0\text{g}}$,
NESCL undergoes isothermal (fixed temperature $T_{0}$) structural relaxation
in time towards ESCL. The entropy of ESCL is shown to extrapolate to zero, but
that of the glass to a non-zero value $S_{\text{R}}>0$ at absolute zero. (b)
The communal entropy of ESCL for a system with an ideal glass transition at
$T_{0\text{K}}$, below which we obtain an ideal glass of zero entropy. }%
\label{Fig_entropyglass-Ideal}%
\end{center}
\end{figure}

The relaxation time $\tau$ of the system usually increases monotonically with
the ratio $\zeta_{0}$; see the upper axis in Fig.
\ref{Fig_Glass_Transition_Time}. Most often, a glass is produced by
supercooling a liquid by \emph{avoiding} crystallization at the melting
temperature $T_{0\text{m}}$; the SCL can still remain under equilibrium in
$\mathcal{D}$ as ESCL shown by the solid curve as long as $\tau<$
$\tau_{\text{exp}}$, but begins to fall out of constrained equilibrium at
$T_{0\text{g}}$ and becomes NESCL as soon as $\tau_{\text{exp}}\simeq\tau$.
This is shown by the dotted portion of the curve in Fig.
\ref{Fig_entropyglass-Ideal}(a), where we show the entropy. The system is not
really frozen for $\tau>>\tau_{\text{exp}}$, but eventually turns into a glass
(GS) at a somewhat lower temperature $T_{0\text{G}}$ (not shown in Fig.
\ref{Fig_Glass_Transition_Time}), where the system will appear to have no
discernible mobility for $\tau>>\tau_{\text{exp}}$. The loss of mobility
results in "freezing" of the system \emph{without} any anomalous changes in
its thermodynamic densities in the glass transition region $(T_{0\text{g}}%
$-$T_{0\text{G}})$. The state below $T_{0\text{G}}$ is identified as a glass.
The two dashed-dotted curves in Fig. \ref{Fig_Glass_Transition_Time}
representing two different glasses will not show any singularity at their
respective glass transition temperature. They smoothly connect with ESCL.
Thus, the \emph{experimental glass transition} should be thought of as a
crossover phenomenon with a gradual turnover of ESCL through NESCL into a GS
over a temperature range.

The EL and ESCL are shown in Fig. \ref{Fig_Glass_Transition_Time} by the
portions of \ the thick solid curve above and below $T_{0\text{m}}$,
respectively, with CR shown by the thin solid curve. There is no abnormal
behavior observed in going from EL into ESCL\ at $T_{0\text{m}}$; contrary to
that, CR properties show a discontinuity \ (see the vertical dashed line at
$T_{0\text{m}}$) with respect to the liquid at $T_{0\text{m}}$%
.\footnote{\label{FN_Singularity}As is well known, this singular behavior
emerges as the system makes a transition between two distinct
states:\ disordere and ordered. The absence of such a singular behavior
between EL and ESCL is because both of them represent the same macrostate. For
the same reason, there is no singular behavior as we go from ESCL to NESCL to
GS. However, the sharp kink $T_{0\text{K}}$ is because IG is a distinct state
different from ESCL. This singularity is coming from the vanishing of the
$V_{\text{f}}$. As the volume $V_{\text{ESCL}}$ of ESCL at $T_{0\text{K}}$
shows no singularity, it can be continued \emph{mathematically }to lower
temperatures, erxcept that $V_{\text{f}}$ will become negative under
mathematical continuation (extrapolation). This is reminiscent of the Kauzmann
paradox.} The relaxation time $\tau$ and viscosity $\eta$ increase by several
orders of magnitudes, typically within a range of a few decades of the
temperature as it is lowered, and eventually surpass experimental limits
$\tau_{\text{exp}}$ and $\eta_{\text{exp}}$, respectively. There is an
"apparent" discontinuity in the slopes over a non-zero temperature range on
the two sides at $T_{0\text{g}}$ as seen in Fig.
\ref{Fig_Glass_Transition_Time}; see also Fig. \ref{Fig_entropyglass-Ideal}%
(a).\footnote{As the curves emerge out of ESCL continuously, there is no
mathematical singularity for the top two curves in Fig.
\ref{Fig_Glass_Transition_Time}; see also footnote \ref{FN_Singularity}.} The
resulting glassy states are represented by the dashed-dotted curves or the
dashed curve. The value of $T_{0\text{g}}$ depends on the external pressure
$P_{0}$ at which the glass former is cooled. One can also obtain a glass
transition by fixing the external temperature and varying the external
pressure. At the corresponding glass transition pressure $P_{0\text{g}}%
(T_{0})$, one will find various densities to have a discontinuity in their
slopes, this time with respect to $P_{0}$.%
%TCIMACRO{\FRAME{ftbpFU}{3.589in}{1.9899in}{0pt}{\Qcb{(a) Schematic behavior of
%the communal entropy of ESCL (solid curve) and a possible time-dependent
%supercooled liquid NESCL, which turns into a glass\ (dotted curve) during
%vitrification. It is assumed that there is no ideal glass transition in ESCL.
%The transition region between $T_{0\text{g}}$ and $T_{0\text{G}}$ has been
%exaggerated to highlight the point that the glass transition is not a sharp
%point.\ For all temperatures $T_{0}<T_{0\text{g}}$, NESCL undergoes isothermal
%(fixed temperature $T_{0}$) structural relaxation in time towards ESCL. The
%entropy of ESCL is shown to extrapolate to zero, but that of the glass to a
%non-zero value $S_{\text{R}}>0$ at absolute zero. (b) The communal entropy of
%ESCL for a system with an ideal glass transition at $T_{0\text{K}}$, below
%which we obtain an ideal glass of zero entropy. }}%
%{\Qlb{Fig_entropyglass-Ideal}}{Graphic3.eps}%
%{\special{ language "Scientific Word";  type "GRAPHIC";
%maintain-aspect-ratio TRUE;  display "USEDEF";  valid_file "F";
%width 3.589in;  height 1.9899in;  depth 0pt;  original-width 4.3889in;
%original-height 2.4517in;  cropleft "-0.0140";  croptop "1";  cropright "1";
%cropbottom "0";  filename 'Graphic3.eps';file-properties "NPEU";}}}%
%BeginExpansion
%EndExpansion

\subsection{Ideal Glass: Analytic Continuation}

The continuous blue curve in Fig. \ref{Fig_entropyglass-Ideal}(a) shows
$S_{\text{ESCL}}$ of ESCL. The red dotted curve show $S_{\text{GS}}$, which
extrapolates to a positive value of the residual entropy $S_{\text{R}}$ at
absolute zero at O as shown. In contrast, it is commonly believed that
$S_{\text{ESCL}}$ extrapolates to zero at $T_{0}=0$ at O, even though there
are no thermodynamic requirements for this

The system shown in Fig. \ref{Fig_entropyglass-Ideal}(a) has no ideal glass
transition. However, the system shown in Fig. \ref{Fig_entropyglass-Ideal}(b),
where we show the communal entropy, undergoes an ideal glass transition at
$T_{0\text{K}}$, where $S^{\text{comm}}$ vanishes. At this point, the system
will undergo a phase transition into IG. From the slope of the entropy, we see
that IG\ has a zero communal heat capacity (heat capacity associated with the
communal entropy), but ESCL has a non-zero communal heat capacity. Thus,
IG\ and ESCL are distinct states, and, as noted in footnote
\ref{FN_Singularity}, the transition will result in a singularity in the
thermodynamic free energy. Despite this singularity, each state itself is
nonsingular. For example, $S_{\text{ESCL}}$ can be mathematically extended to
temperatures below $T_{0\text{K}}$, although it will result in a negative
communal entropy. This is clearly seen in the Gibbs-Di Marzio theory, where
the free energy can be mathematically continued all the way down to absolute zero.

\subsection{Real Glass and Approximate Thermodynamics}

Theoretical and experimental investigations of laboratory glasses invariably
but not always require applying \emph{equilibrium} (time-independent)
thermodynamics to SCL to extract quantities like the entropy. This
approximation is equivalent to treating SCL as ESCL obtained under infinitely
slow cooling$.$ It is further assumed that the SCL free energy can be defined
all the way down to absolute zero, as was the case with the Gibbs-Di Marzio
theory. This may not always be possible as the extension may terminate in a
\emph{spinodal} at a non-zero temperature as we lower the temperature.

\section{Nonequilibrium Formulation:\ Brief
Review\label{Nonequilibrium_Thermodynamics}}

We consider an interacting system $\Sigma$ embedded in a medium $\widetilde
{\Sigma}$; their combination forms an isolated system $\Sigma_{0}$. Quantities
pertaining to $\Sigma_{0}$ will have a suffix $0$, while those for
$\widetilde{\Sigma}$ will have a tilde; quantities for $\Sigma$ will have no
suffix. The medium is taken to be extremely large compared to $\Sigma$ so that
the latter does not affect the fields of $\widetilde{\Sigma}$, which can be
taken to be equal to that of $\Sigma_{0}$ so that we will be denoted by
$T_{0},P_{0},$ etc. We will also find it convenient to use body to denote any
one of $\Sigma,$ $\widetilde{\Sigma}$ and $\Sigma_{0}$. The quantities
pertaining to a body will not have any suffix.

\subsection{Equilibrium and Nonequilibrium States}

Let $\Sigma$ be a nonequilibrium state such as a glass. glass is a
nonequilibrium microstate, its entropy $S(t)$ must obey the second law, i.e.,
the law of increase of entropy, according to which the \emph{irreversible}
(denoted by a suffix i)\footnote{This entropy is generated within the system
because of dissipative processes. Thus the suffix i can also stand for
"internal." The quantity $d_{\text{e}}S$ with a suffix "e" will denote the
entropy exchange with the medium (the outside).} entropy generated in any
infinitesimal physical process going on within a body satisfies the inequality%
\begin{equation}
d_{\text{i}}S\geq0; \label{Second_Law0}%
\end{equation}
the equality $d_{\text{i}}S=0$ occurs for a reversible process. For an
isolated system $\Sigma_{0}$, there is no \emph{exchange} (denoted by a suffix
e) entropy change $d_{\text{e}}S_{0}=0$ so that in any arbitrary process
satisfies%
\begin{equation}
dS_{0}=d_{\text{i}}S_{0}\geq0. \label{Second_Law}%
\end{equation}
For $\Sigma_{0}$ in equilibrium, $dS_{0}=0$ so that its entropy is constant.

\subsection{Concept of a Nonequilibrium State and of\ Internal
Equilibrium\label{Marker_Internal Equilibrium}}

\subsubsection{Isolated Body}

For $\Sigma_{0}$ in equilibrium, $S_{0}$ can be expressed as a \emph{state
function} $S_{0}(\mathbf{X}_{0})$\ of its extensive observables (variables
such as energy, volume, particle number, etc. that can be controlled by an
observer) $\mathbf{X}_{0}$, which remain constant. The thermodynamic state of
a body in equilibrium remains the same unless it is disturbed. From this
follows the Gibbs fundamental relation
\begin{equation}
dS_{0}=%
%TCIMACRO{\tsum \nolimits_{p}}%
%BeginExpansion
{\textstyle\sum\nolimits_{p}}
%EndExpansion
\left(  \partial S_{0}/\partial X_{0p}\right)  dX_{0p},
\label{Gibbs_Fundamental}%
\end{equation}
in which the partial derivatives are related to the constant \emph{fields} of
the system:%
\begin{equation}
\left(  \partial S_{0}/\partial E_{0}\right)  =1/T_{0},\left(  \partial
S_{0}/\partial V_{0}\right)  =P_{0}/T_{0},\cdots\label{Fields_Isolated}%
\end{equation}

For $\Sigma_{0}$ out of equilibrium, its nonequilibrium state will
continuously change, which is reflected in its entropy increase in time. This
requires expressing its entropy as $S_{0}(\mathbf{X}_{0},t)$ with an
\emph{explicit time-dependence}. The change in the entropy and the state must
come from the variations of additional variables, distinct from $\mathbf{X}%
_{0}$, that keep changing with time until the body comes to equilibrium
\cite{Gujrati-I,Nemilov}. These variables, whose set is denoted by
$\boldsymbol{\xi}_{0}$, cannot be controlled by the observer; thus, they are
known as the \emph{internal variables}. Once $\Sigma_{0}$ has come to
equilibrium, $S_{0}$ has no explicit time-dependence and becomes a state
function of $\mathbf{X}_{0}$, which requires that $\boldsymbol{\xi}_{0}$ are
no longer independent of $\mathbf{X}_{0}$. When $S_{0}$ becomes a state
function, it achieves its maximum possible value for the given set
$\mathbf{X}_{0}$. This conclusion about the entropy will play an important
role below.

We will refer to the variables in $\mathbf{X}_{0}$ and $\boldsymbol{\xi}_{0}%
$\ collectively as $\mathbf{Z}_{0}$ in the following. It can be shown that
with a proper choice of the number of internal variables, the entropy can be
written as $S_{0}(\mathbf{Z}_{0}(t))$ with \emph{no} explicit $t$-dependence.
The situation is now almost identical to that of an isolated body in
equilibrium:\ As the entropy $S_{0}(\mathbf{Z}_{0}(t))$ no longer has an
explicit time dependence, we can identify $\mathbf{Z}_{0}(t)$\ as the set of
\emph{nonequilibrium state variables} so that its state can be specified by
$\mathbf{Z}_{0}(t)$, which makes the entropy a state function. From now on, we
will not show the time $t$ as an argument of a state variable or a state
function unless clarity is needed. There are times we will insert $t$Thus, we
can extend (\ref{Gibbs_Fundamental}) to%
\begin{equation}
dS_{0}=%
%TCIMACRO{\tsum \nolimits_{p}}%
%BeginExpansion
{\textstyle\sum\nolimits_{p}}
%EndExpansion
\left(  \partial S_{0}/\partial Z_{0p}\right)  dZ_{0p};
\label{Gibbs_Fundamental_Extended}%
\end{equation}
the new derivatives $\left(  \partial S_{0}/\partial\xi_{0p}\right)  $ (in
addition to those in (\ref{Fields_Isolated})) determine the \emph{affinities}
$A_{0p}$%
\begin{equation}
\left(  \partial S_{0}/\partial\xi_{0p}\right)  =A_{0p}/T_{0};
\label{Affinities_Isolated}%
\end{equation}
all fields and affinities continue to change in time until $\Sigma_{0}$
reaches equilibrium. An isolated body for which the entropy has become a state
function of $\mathbf{Z}_{0}$is called to be in \emph{internal equilibrium}. As
the body comes to equilibrium, the affinities $A_{0p}$vanish. Moreover,
$S_{0}$ as a state function has its maximum possible value for given
$\mathbf{Z}_{0}$. For a state that is not in internal equilibrium, its $S_{0}$
must retain an explicit time-dependence. In this case, the derivatives in
(\ref{Fields_Isolated}) \emph{cannot} be identified as state variables like,
temperature, pressure, etc.

\subsubsection{A Simple Example for an Internal Variable}

Consider an isolated body $\Sigma_{0}$ formed by two parts $\Sigma_{1}%
,\Sigma_{2}$ that are initially at different temperatures $T_{1}(0)$ and
$T_{2}(0)$. We imagine their volumes and particle numbers as fixed, but allow
their energies $E_{1},E_{2}$to change with time due to thermal contact between
them. We know that eventually, they come to equilibrium at the same common
temperature $T_{0\text{f}}$. During this time, their temperatures keep
changing. The total energy $E_{0}=E_{1}+E_{2}$ is constant. The entropy
$S_{0}$ is the sum of the entropies $S_{1}$ and $S_{2}$ of the two parts.
Thus, $S_{0}$ is a function of two variables $E_{1},E_{2}$. If we want to
express $S_{0}$ as a function of $E_{0}$, we need another\emph{ independent}
variable $\xi_{0}$, which is evidently a function of the difference
$E_{1}-E_{2}$. We take $\xi_{0}\equiv\lbrack E_{1}-E_{2}]/2$. This makes
$S_{0}$ a function of $E_{0}$ and $\xi_{0}.$ The affinity $A$is given by the
derivative $\partial S_{0}/\partial\xi_{0}$, which is found to be
$A=1/T_{1}-1/T_{2}$and which vanishes in equilibrium,as expected. Here,
$T_{1}$and $T_{2}$are the instantaneous temperatures of the two parts and are
given by $1/T_{1}=\left(  \partial S_{1}/\partial E_{1}\right)  ,1/T_{2}%
=\left(  \partial S_{2}/\partial E_{2}\right)  $.

\subsubsection{Interacting Body}

An interacting body (a body in a medium) out of equilibrium with its medium
will also require internal variables. For a body in internal equilibrium, the
derivatives of its entropy $S\ $give the fields $T,P,\cdots$ and affinities
$A_{p}$as above. The corresponding Gibbs fundamental relation is given by%
\[
dE=TdS-PdV-\mathbf{A}\cdot d\mathbf{\xi},
\]
where we have restricted the observables to $E$ and $V$ only for simplicity.

\subsubsection{Why Internal Equilibrium is Important?}

For a general interacting body, the concept of its internal equilibrium state
plays a very important role in that it ensures that the body can come back to
this state several times in a nonequilibrium process. This can only happen if
the entropy have no explicit time-dependence. In a cyclic nonequilibrium
process, such a state can repeat itself in time after some cycle time
$\tau_{\text{c}}$ so that all state variables and functions including the
entropy repeat themselves:$\mathbf{Z}(t+\tau_{\text{c}})=\mathbf{Z(}%
t\mathbf{)},~S(t+\tau_{\text{c}})=S(t).$

\subsection{Gibbs Free Energy of an Interacting
System\label{Marker_Gibbs_Free_Energy}}

From now on, we will only consider bodies in internal equilibrium. We will
further simplify our discussion by considering only one internal variable
$\xi$ in most cases. Moreover, we will keep $N$ fixed so that it will not be
shown explicitly and allow the possibility of fluctuating $E$ and $V$ due to
exchanges with the medium. The medium is in internal equilibrium and is
extremely large compared $\Sigma$ so that its fields and affinity are given by
$\ T_{0},P_{0}$ and $A_{0\xi}=0$, the values they take when the equilibrium is
reached between $\Sigma$ and $\widetilde{\Sigma}$. In terms of
\begin{align*}
H  &  \equiv E+P_{0}V,\ G\equiv H-T_{0}S\\
&  =E-T_{0}S+P_{0}V,
\end{align*}
which are the time-dependent enthalpy and the Gibbs free energy, respectively,
of the system $\Sigma$, it is easy to show that \cite{Gujrati-I}
\begin{equation}
S_{0}-\widetilde{S}_{0}=S-H/T_{0}=-G/T_{0},
\label{Gibbs_Free_Energy_Entropy_Relation}%
\end{equation}
where $\widetilde{S}_{0}\equiv\widetilde{S}(E_{0},V_{0},\xi_{0})$ is
independent of the system. Here, $S$ and $S_{0}$are the entropies of $\Sigma$
and $\Sigma_{0}$.

\section{Nonequilibrium Relaxation of Bodies in Internal Equilibrium}

\subsection{Thermodynamic Relaxation}

\subsubsection{Heuristic Consideration}

Let us consider $\Sigma$, originally in equilibrium with a medium at some
temperature $T_{0}^{\prime}$, is suddenly brought in another medium at a lower
temperature $T_{0}$ at time $t=0$. As $\Sigma$ has no time to interact with
the new medium, its initial temperature $T(0)$ is the original temperature
$T_{0}^{\prime}$. As $\Sigma$ eventually comes to equilibrium, we must have
$T(\tau_{\text{eq}}(T_{0}))=T_{0}$, where $\tau_{\text{eq}}(T_{0})$\ is the
time required to come to equilibrium. Thus, $T$continues to fall during
relaxation. Similarly, the initial entropy\ $S(0)$ is the entropy
$S_{\text{ESCL}}(T_{0}^{\prime})$ of the higher temperature. After
equilibration, $S(T_{0}^{\prime})S_{\text{ESCL}}(T_{0})$the entropy must be
the entropy of ESCL at $T_{0}$ at the new temperature. Since the entropy
increases with temperature, we conclude that the entropy also falls during
relaxation from $S(\tau_{\text{eq}}(T_{0}))=S(T_{0})$.

\subsubsection{Thermodynamic Support}

We now support this intuitive picture for the temperature by proper
nonequilibrium thermodynamics. Here, we closely follow Ref. \cite{Gujrati-I}.
The Gibbs fundamental relation for the system (fixed $N$) is given by%
\begin{equation}
dE=TdS-PdV-Ad\xi. \label{Gibbs_Fundamental_Equation00}%
\end{equation}
The internal fields $T,P$and $A$are given, respectively by the derivatives
$(\partial E/\partial S),-(\partial E/\partial V)$ and $-(\partial
E/\partial\xi)$. It can be shown that%

\begin{align}
d_{\text{i}}S/dt  &  \equiv dS_{0}/dt\nonumber\\
=\left(  1/T-1/T_{0}\right)  dE/dt  &  +\left(  P/T-P_{0}/T_{0}\right)
dV/dt\label{Total_Entropy_Rate}\\
+[A/T]d\xi/dt  &  \geq0.\nonumber
\end{align}
Each term in the last equation must be positive in accordance with the second
law. In a vitrification process, energy decreases with time; thus,%
\begin{align}
T  &  >T_{0}\text{ if }dE/dt<0,T<T_{0}\text{ if }dE/dt>0;\nonumber\\
P/T  &  <P_{0}/T_{0}\text{ if }dV/dt<0,P/T>P_{0}/T_{0}\text{ if }%
dV/dt>0,\label{Field_Behavior}\\
A/T  &  <0\text{ if }d\xi/dt<0,A/T>0\text{ if }d\xi/dt>0.\nonumber
\end{align}

Let us assume that in an \emph{isobaric} cooling experiment,%
\begin{equation}
P=P_{0}, \label{Mech_Equil}%
\end{equation}
which we will refer to as the existence of the \emph{mechanical equilibrium
}for the system. In this case, we find by combining the first two terms in
(\ref{Total_Entropy_Rate}) that
\begin{equation}
\left(  1/T-1/T_{0}\right)  dH/dt\geq0. \label{Isolated_Entropy_Variation}%
\end{equation}
It is found experimentally \cite{Goldstein}
\begin{equation}
T>T_{0} \label{Temperature_Behavior}%
\end{equation}
during relaxation in glasses so that $T$approaches $T_{0}$ from above
[$T\rightarrow$ $T_{0}^{+}$] and becomes equal to $T_{0}$ as the relaxation
ceases and the equilibrium is achieved.

\subsection{Thermodynamic Forces for Relaxation}

For $P\neq P_{0}$, the Gibbs fundamental relation can be written as
\begin{align}
dE  &  =T_{0}dS-P_{0}dV\nonumber\\
+(T-T_{0})dS  &  -(P-P_{0})dV-Ad\xi,
\label{Gibbs_Fundamental_Equation_Expanded00}%
\end{align}
in which each of the last three terms are associated with an irreversible
entropy generation \cite{Gujrati-I}. Using the first law $dE=T_{0}d_{\text{e}%
}S-P_{0}dV$, we find that the irreversible entropy change $d_{\text{i}}S$is
\ \ \ \ \
\[
T_{0}d_{\text{i}}S=(T_{0}-T)dS+(P-P_{0})dV+Ad\xi\geq0.
\]
Each of the three terms on the right must be non-negative in accordance with
the second law
\begin{equation}
(T_{0}-T)dS\geq0,(P-P_{0})dV\geq0,Ad\xi\geq0.
\label{Second_Law_Inequalities00}%
\end{equation}
The prefactor $T_{0}-T$, etc. in each equation represents the
\emph{thermodynamic force} that drive the system towards equilibrium. It then
follows from the first inequality in (\ref{Second_Law_Inequalities00}) that
during vitrification%
\begin{equation}
dS/dt\leq0, \label{Entropy_variation}%
\end{equation}
in accordance with our heuristic consideration.

\subsection{Consequences of the Customary Approximation}

For the isobaric case discussed above, we have assumed $P=P_{0}$, but
$T_{0}-T$is normally non-zero. However, in almost all previous applications of
classical non-equilibrium thermodynamics to glasses that we are familiar with,
$T_{0}-T$is taken to be zero \cite{Nemilov}. It is important to understand the
consequence of this customary approximation for glasses. We have in this
case,\ $T_{0}d_{\text{i}}S=Ad\xi\geq0$, from which follows that $T_{0}%
dS=T_{0}d_{\text{e}}S+Ad\xi$. As $d_{\text{e}}S<0$ in cooling, there is no way
to make any conclusion about the sign of $dS$: it may be even positive. From
(\ref{Isolated_Entropy_Variation}) and
(\ref{Gibbs_Free_Energy_Entropy_Relation}), we see neither the Gibbs free
energy $G$ of the glass nor the entropy $S_{0}$ of the isolated system can
vary in time. Thus, there will be no relaxation of the Gibbs free energy even
if we allow for internal variables.

\subsection{Microstate Probabilities in Internal Equilibrium}

For a body in internal equilibrium, the situation is very similar to that of a
body in equilibrium in that maximization of entropy results is a very similar
formulation of the microstate probabilities. While these probabilities are
given by%
\[
p_{i\text{eq}}=Q_{0\text{eq}}\exp[-\beta_{0}(E_{i}+P_{0}V_{i})],
\]
where $Q_{0\text{eq}}$ is the equilibrium normalization constant, $\beta
_{0}=1/T_{0}$ and $P_{0}$\ are the inverse temperature and pressure of the
medium, and $E_{i},V_{i}$ are the energy and volume of the $i$th microstate.
For the body in internal equilibrium, the microstate probabilities are given
by
\[
p_{i}=Q_{0}\exp[-\beta\{E_{i}+PV_{i}+\mathbf{A}\cdot\mathbf{\xi}_{i}\}],
\]
where $Q_{0}$ is the normalization constant; $\beta=1/T$, $T,P,\mathbf{A}$the
instantaneous temperature, pressure and the set of affinities, and
$\mathbf{\xi}_{i}$\ the set of internal variables for the $i$th microstate.
The particle number $N$, which is held fixed in both cases, is not shown.

\section{Free Volume and Communal Entropy:\ Cell and Hole Theories}

In this section, we will exhibit $t$ as an argument if clarity is needed.

\subsection{The Cell Theory}

\begin{figure}
[ptb]
\begin{center}
\includegraphics[
trim=-0.123192in 0.000000in -0.123192in 0.000000in,
height=1.8118in,
width=3.3987in
]%
{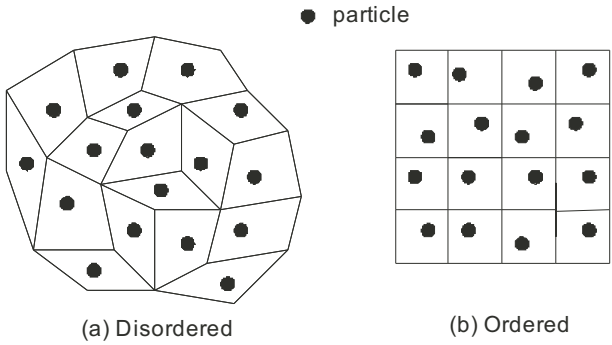}%
\caption{Cell representation of a small region of disordered (a) and ordered
(b) configurations at full occupation: each cell contains a particle. Each
cell representation uniquely defines a potential well or basin in the
potential energy landscape. Observe that while each particle is surrounded by
four particles in the ordered configuration, this is not the case for the
disordered configuration. We have shown a higher volume for the disordered
configuration, as found empirically. }%
\label{Fig_Random_Ordered_States0}%
\end{center}
\end{figure}

One usually studies a system by treating the constituent particles to be point
like such as the ideal gas. The approximation allows us to think of the entire
volume $V$ of the system as the "free volume" in which the (center of mass of
the) particles are \emph{free} to move about. The simplest model to account
for the non-zero size of the particles is the van der Waals' equation in which
the "free volume" is given by $V-Nb$, where $b$ is taken to be half the the
volume of a sphere of radius $2r_{0}$, $r_{0}$ being the "radius" of the
particle \cite{Landau}. It is the excluded volume for each particle and the
presence of $N$ in $Nb$ implies the additivity of the excluded volume per
particle. The excluded volume should be thought of \ as the "thermodynamic"
volume of a particle, which is determined by the interactions with its
neighbor. One uses the cell theory of liquids to go beyond the van der Walls
theory. In the cell model, the volume $V$ is divided into $N$ cells (such as
the Voronoi type cells) of an average volume $v$ per particle, see Fig.
\ref{Fig_Random_Ordered_States0}, where we show the possible cell arrangement
for disordered (liquid or gas) in (a) and ordered (crystal) states in (b).
Each cell has a single particle within it as shown. For molecules with
connectivity such as polymers, one must take proper care of all
\emph{distinct} placements of monomers that respects their connectivity. For
example, if we consider a disordered conformation of a polymer with $17$
monomers, then we must consider all distinct conformations of the polymer even
though each conformation has a single monomer in the $17$ cells in.(a). Thus,
there will be many more microstates for the cell pattern in (a) when
connectivity has to be incorporated. There will be a single microstate if
there is no connectivity to consider. This poses no conceptual problem as the
average of any observable $O$ in the cell model is given by the standard
formulation%
\[
\overline{O}\equiv%
%TCIMACRO{\tsum \nolimits_{i}}%
%BeginExpansion
{\textstyle\sum\nolimits_{i}}
%EndExpansion
O_{i}p_{i},
\]
where $O_{i}$ is the value of $O$ for the $i$th microstate and the sum is over
all distinct microstates. It is then clear that the entropy associated with
local motion in the cell potential also contains what is commonly known as the
\emph{conformational entropy} due to different conformations of a polymer.%

%TCIMACRO{\FRAME{ftbpFU}{3.3987in}{1.8118in}{0pt}{\Qcb{Cell representation of a
%small region of disordered (a) and ordered (b) configurations at full
%occupation: each cell contains a particle. Each cell representation uniquely
%defines a potential well or basin in the potential energy landscape. Observe
%that while each particle is surrounded by four particles in the ordered
%configuration, this is not the case for the disordered configuration. We have
%shown a higher volume for the disordered configuration, as found empirically.
%}}{\Qlb{Fig_Random_Ordered_States0}}{Random_Graph0.eps}%
%{\special{ language "Scientific Word";  type "GRAPHIC";
%maintain-aspect-ratio TRUE;  display "USEDEF";  valid_file "F";
%width 3.3987in;  height 1.8118in;  depth 0pt;  original-width 4.248in;
%original-height 2.3782in;  cropleft "-0.0290";  croptop "1";
%cropright "1.0290";  cropbottom "0";
%filename 'Random_Graph0.eps';file-properties "NPEU";}}}%
%BeginExpansion
%EndExpansion

The non-uniform cell model in (a) for the disordered state is a generalization
of the uniform cell model traditionally used in cell theories. The other
difference is that we allow for an internal variable. The motion of each
particle within its cell is governed by $\varphi$ due to all its neighbor,
which we denote by $\varphi(\mathbf{r}\left\vert \{\mathbf{r}^{\prime
}(t)\}\right.  )$, where $\mathbf{r}$ is the position of the particle under
consideration within its cell and $\{\mathbf{r}^{\prime}\}$ denotes the set of
the positions of all its neighbors that are continually c hanging in time. The
connectivity between the neighbors and the central molecule has to be properly
accounted for in this set. This requires considering the probability
$P(\varphi)$ for the potential $\varphi(\mathbf{r}\left\vert \{\mathbf{r}%
^{\prime}(t)\}\right.  )$. It is given by the following identity%
\[
P(\varphi)=%
%TCIMACRO{\tsum \nolimits_{i}^{^{\prime}}}%
%BeginExpansion
{\textstyle\sum\nolimits_{i}^{^{\prime}}}
%EndExpansion
p_{i}%
\]
where the prime over the sum implies that the sum is restricted to those
microstates in which the particle and its cell neighbors are restricted to be
at $\mathbf{r}$ and $\mathbf{r}^{\prime}$.

It is obvious that the cell volume $\Delta$ must be at least as big as needed
to allow for the local oscillations. The local motion within the cell
potential can be used to characterize a volume, which we will call the
\emph{interaction volume} $v_{\text{i}}$ of the particle: it is the minimum
volume needed for the allowed local motion. One way to quantify $v_{\text{i}}$
is by the root-mean square-root of the displacement during the local motion as
follows. The average over all possible cell potentials with probability
$P(\varphi)$ of the displacement squared is given by
\[
r_{\text{rms}}^{2}\equiv\overline{\mathbf{r}^{2}}=%
%TCIMACRO{\tsum \nolimits_{\varphi}}%
%BeginExpansion
{\textstyle\sum\nolimits_{\varphi}}
%EndExpansion
P(\varphi)[\frac{1}{\tau_{\varphi}}%
%TCIMACRO{\tint }%
%BeginExpansion
{\textstyle\int}
%EndExpansion
\mathbf{r}^{2}(t)dt],
\]
where $\mathbf{r}=0$ is taken to be at the minimum of the potential in the
inherent structure IS. The inner integral is over the time period
$\tau_{\varphi}$ of an oscillation controlled by $\varphi$. We first observe
that the $r_{\text{rms}}$ changes with the temperature. To see this most
clearly, we simply consider an interparticle harmonic potential $\varphi$.
From dimensional analysis alone, we conclude that the mean square displacement
of all the particles scales as the temperature:$\overline{r^{2}}\sim T(t)/k$,
where $k$ is the spring constant of the potential. Thus,%
\[
r_{\text{rms}}(T(t)\sim\left[  T(t)/k\right]  ^{1/2}.
\]
Accordingly, it vanishes at absolute temperature $T(t)=0$. It usually happens
that the oscillatory modes equilibrate rapidly with the medium. In that case,
we must replace $T(t)$ by $T_{0}$. At absolute zero, particles are sitting in
an IS, and there is an average distance $d_{\text{min}}$ between particles.
Half of the average distance $r_{\text{min}}\equiv d_{\text{min}}/2$ is taken
as the "radius" of the particle, which then determines its interaction volume
$v_{\text{m}}$ at absolute zero. This is the minimum of the interaction
volume. The distance $r_{\text{min}}$ should not be confused with the
so-called radius $r_{0}$ of a particle corresponding to the "impenetrability"
of the particles.\footnote{The interparticle potential begins to rise steeply
for separation between particles below the "radius" $r_{0}$ \cite{Landau}.
Thus, $r_{0}$ not strictly the radius of a particle.} As the temperature
increases, the linear size of the particle increases due to oscillations and
so does its interaction volume, which is now given by%
\[
v_{\text{i}}=\gamma(r_{\text{min}}+r_{\text{rms}})^{3}%
\]
in terms of a geometrical factor $\gamma$ of order unity. The above volume may
be quite different from the customarily defined occupied volume $v_{\text{o}}%
$, which is commonly used in the glassy literature. Unfortunately, there is a
lot of ambiguity in the definition of $v_{\text{o}}$ and how to obtain it
theoretically \cite{Matsuoka} so we make no attempt to compare the two.

The difference $v_{\text{f}}=\Delta-v_{\text{i}}$ is the remainder volume of
the cell, called the free volume $v_{\text{f}}$, that is the elbow room for
the translation and rotation of the center of mass of the particle. This
motion gives rise to the diffusion of the particle from the region over which
the local oscillatory motion occurs. As $\zeta_{0}$ decreases, the elbow room,
i.e., $v_{\text{f}}$ increases (and so do $\Delta$ and $v_{\text{i}}$) but not
so much so that particles are still confined within their respective cells. As
the free volume increases further, the particles can escape to the neighboring
cells so that sometimes a cell may have multiple occupancy of the particles.
The particles will undergo local motion in the new cell before they make
excursion to another neighboring cell. If the free volume increases too much,
then diffusion becomes the dominant motion and the local motion is no longer
possible as the particles are far apart now. A situation like this occurs in
gases. The above picture is an average picture so that it will also occur due
to fluctuations in energy,volume, and internal variables. The aforementioned
scenario has been confirmed by numerical simulations that has been discussed
by several authors; see for example, Refs. \cite{Zallen,Gujrati-I} and Fig.
\ref{Fig_Simulation}.

The above discussion is equally valid for the disordered and ordered
macrostate. We will, however, consider the disordered macrostate in the following.

\subsection{The Hole Theory}

\begin{figure}
[ptb]
\begin{center}
\includegraphics[
trim=-0.123137in 0.000000in -0.123137in 0.000000in,
height=2.0851in,
width=3.4195in
]%
{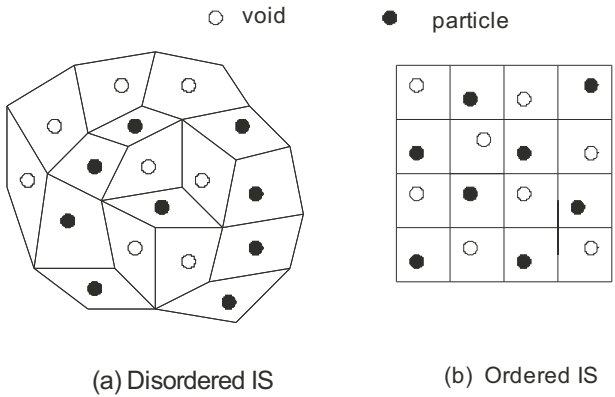}%
\caption{Cell representation of a small region of disordered (a) and ordered
(b) inherent structures at half occupation: half of the cells contain a
particle; other half are empty and are said to contain a void. Observe that
while each particle is surrounded by four voids in the ordered configuration,
this is not the case for the disordered configuration. We have shown a higher
volume for the disordered configuration, as found empirically.}%
\label{Fig_Random_Ordered}%
\end{center}
\end{figure}

An obvious refinement of the above cell theory is to allow for holes in the
theory; see (a) and (b) in Fig. \ref{Fig_Random_Ordered}. The empty sites
refer to holes, that is, the absence of a particle and gives an additional
contribution to the free volume in this theory. It also allows for a
considerable variation in the number of neighboring particles due to the
presence of holes, which makes this theory attractive. As the volume of a
glass is found to be considerably greater than that of the corresponding
crystal or ESCL, it is argued that the glass has a significant number of
holes, which decreases during relaxation. In general, the decrease in the free
volume is considered to be related to the irreversible relaxation, while the
decrease in the interaction volume is argued to be related to the equilibrium
relaxation as noted above since the local motion within the cell potential
occurs at the temperature $T_{0}$ of the medium and not at the instantaneous
temperature of the glass; see also the discussion by Matsuoka \cite{Matsuoka}.
The division of volume in the interaction and free volumes results in the two
volumes to be independent as they refer to independent degrees of freedom.
Their existence may be related to the success of the two parameter model of
Aklonis and Kovacs \cite{Kovacs}.%
%TCIMACRO{\FRAME{ftbpFU}{3.4195in}{2.0851in}{0pt}{\Qcb{Cell representation of a
%small region of disordered (a) and ordered (b) inherent structures at half
%occupation: half of the cells contain a particle; other half are empty and are
%said to contain a void. Observe that while each particle is surrounded by four
%voids in the ordered configuration, this is not the case for the disordered
%configuration. We have shown a higher volume for the disordered configuration,
%as found empirically.}}{\Qlb{Fig_Random_Ordered}}{Random_Graph.eps}%
%{\special{ language "Scientific Word";  type "GRAPHIC";
%maintain-aspect-ratio TRUE;  display "USEDEF";  valid_file "F";
%width 3.4195in;  height 2.0851in;  depth 0pt;  original-width 4.2756in;
%original-height 2.7423in;  cropleft "-0.0288";  croptop "1";
%cropright "1.0288";  cropbottom "0";
%filename 'Random_Graph.eps';file-properties "NPEU";}}}%
%BeginExpansion
%EndExpansion

\subsection{Communal Entropy\label{Sec._communal_entropy}}

Using the fact that entropy and volume are both extensive, we find that the
entropy density $\sigma=S/V$ is a \emph{homogeneous function} of order zero,
that is, it is not extensive. We can write the entropy as a sum of two terms:
$S=V_{\text{i}}\sigma+V_{\text{f}}\sigma.$ This is a trivial identity but
allows us to introduce two different entropy components associated with the
two components of the volume. One such component is $S^{\text{comm}}$, which
is associated with the translation of the particles. The free volume decreases
as $\zeta_{0}$ increases and inhibits the translation. The particles must be
fully jammed if there is no elbow room ($V_{\text{f}}=0$). In this state,
there cannot be any translation and $S^{\text{comm}}$ must also vanish. This
identifies the \emph{ideal glass} (IG). If we wish to identify the communal
entropy associated with the free volume, then it must vanish for the ideal
glass. Thus, the relationship between $S^{\text{comm}}$ and $V_{\text{f}}$
must be \emph{linear }because of the extensivity. If we write it as
$S^{\text{comm}}=V_{\text{f}}\sigma^{\prime}$ with $\sigma^{\prime}\neq\sigma
$, then the other component $S^{\text{int}}=S-S^{\text{comm}}$, the
interaction entropy, is given by $S^{\text{int}}=V_{\text{i}}\sigma
+V_{\text{f}}(\sigma-\sigma^{\prime})$. However, $S^{\text{int}}$ must be
determined by the cell potential $\varphi$, which depends on $\Delta$ or $V$
but is independent of the free volume. Hence, $\sigma^{\prime}=\sigma$. Thus,
we write
\[
S(t)=S^{\text{int}}(t)+S^{\text{comm}}(t),S^{\text{comm}}=V_{\text{f}}%
\sigma,S^{\text{int}}=V_{\text{i}}\sigma;
\]
each of the above two components of the entropy must be non-negative and must
satisfy the second law. As said earlier, $S^{\text{int}}$ includes the entropy
associated with different conformations of \ the molecules. The communal
entropy $S^{\text{comm}}$ that plays a central role in the study of glasses
\cite{Gujrati-Book,Cohen-Grest,Zallen}.\ The deep connection that we have
discovered between the free volume and communal entropy shows that they vanish
simultaneously in IG so that whether we vary the density (control variable
$P_{0}$) or the entropy (control variable $T_{0}$), we obtain the unique IG at
the respective Kauzmann point. This, we hope, will clarify some confusion
present in the field as we discuss now.

\section{Some Glass Transition Theories}

We briefly review some important theories that have been used to explain glass
transitions; for more details, see Ref. \cite{Gujrati-Book}. None of them at
present is able to explain all observed features of glass transition
\cite{Debenedetti,Debenedetti1}. Thus, we are far from having a complete
understanding of the phenomenon of glass transition, and it is fair to say
that there yet exists no completely satisfying theory of the glass transition.
Theoretical investigations mainly utilize two different approaches, which are
based either on thermodynamic or on kinetic ideas, neither of which seems
complete.%
\begin{figure}
[ptb]
\begin{center}
\includegraphics[
trim=0.000000in 0.798196in 1.498057in 0.000000in,
height=1.9095in,
width=3.6115in
]%
{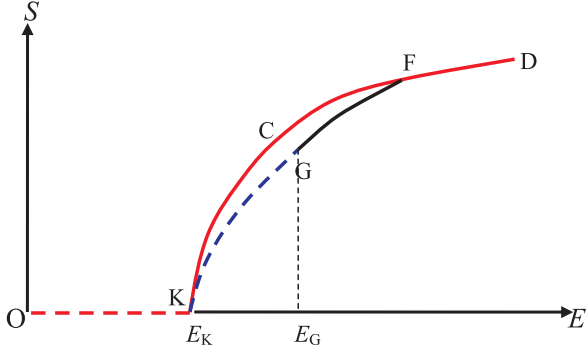}%
\caption{Schematic form of communal entropies for ESCL\ and NESCL in
accordance with the second law and the gap hypothesis. The lowest possible
energy of ESCL is $E_{\text{K}}$ and the corresponding entropy is given by
KCFD. It vanishes at K without any singularity there. The point K represents
the unique ideal glass whose entropy remains zero as shown by the ideal glass
entropy OK. The lowest possible energy of NESCL, which emerges continuously
out of ESCL at F, is shown to be $E_{\text{G}}>E_{\text{K}}$ at G due to
additional defects. It represents the entropy of a laboratory glass and has no
singularity at G so that it can be mathematically continued below
$E_{\text{G}}$ until it vanishes. As the state of zero entropy must be unique,
this continuation must terminate at K where the ideal glass emerges.
Accordingly, the continuation is shown by the dashed blue curve GK. It is in
essence similar to the continuation carried out by Kauzmann. We do not show
the axis corresponding to the independent variable $\xi$ (except at K and F,
where it is no longer independent), which is changing along FG. The slope at G
is less than that at K.}%
\label{Fig.NE-E_SCL._entropy}%
\end{center}
\end{figure}
%EndExpansion

\subsection{Thermodynamic Approach:\ General
Considerations\label{marker_Sec_Thermodynamic Considerations}}

We first focus on our nonequilibrium approach to see what general conclusions
can be drawn before considering other theories. In the following, we are
interested in SCL state, which can be divided into two: time-dependent NESCL
and time-independent ESCL in the restricted ensemble. We will focus on the
communal entropy for which we use $S$ instead of $S^{\text{comm}}$ for
notational simplicity. We will exhibit $E$ and suppress all other extensive
observables below and use a single internal variable $\xi$ for simplicity.
According to the second law $S_{\text{NESCL}}(E,\xi)<S_{\text{ESCL}}(E)$; see
Fig. \ref{Fig.NE-E_SCL._entropy}. As the slope of FG determines the inverse of
the internal temperature $T(t)\neq T_{0}$ of NESCL, while that of DFCK the
inverse of the medium temperature $T_{0}$, we conclude from the figure that
$T_{\text{G}}(t)$ at $E_{\text{G}}$ is higher than $T_{\text{K}}=T_{0\text{K}%
}$ at $E_{\text{K}}$. Indeed, as $T_{\text{NESCL}}(E)=T_{\text{ESCL}}(E)$ at
$E=E_{\text{F}}$ and at $E=E_{\text{K}}$, where $E_{\text{F}}$ is the energy
at F, it is evident that $T_{\text{NESCL}}(t)>T_{0\text{K}}$ over FG. As
$S_{\text{NESCL}}(E,V,\xi)$ has no singularity over its entire range KGF, we
can expand $S_{\text{NESCL}}(E,V,\xi)$ in the form of a Taylor expansion over
FG around the point K. For later applicability, it is useful to consider
$S_{\text{NESCL}}(E,V,\xi)$ as a function of $T(t),P(t)$ and $\xi(t)$. We will
suppress $t$ in the following. In an isobaric vitrification at medium pressure
$P_{0}$, it is not hard to conceive that $P=P_{0}$, which we will assume in
the following. Introducing $\Delta T\equiv T-T_{0\text{K}}$ and $\Delta
\xi\equiv\xi-\xi_{\text{eq,K}}$ and recognizing that at K the heat capacity
$C_{P\text{K}}$ is non-zero but finite and that the affinity $A_{0\text{K}}$
vanishes, we immediately conclude that the leading terms in the expansion must
be linear in $\Delta T$(no linear term in $\Delta\xi$) and bilinear in $\Delta
T\Delta\xi$. Thus, we can pull out $\Delta T$ from all the expansion terms to
finally write in terms of a function $F_{V}$ whose definition is evident from
the (infinite) Taylor expansion ($S_{\text{ESCL}}(E_{\text{K}})=0$):
\begin{equation}
S_{\text{NESCL}}(T,P_{0},\xi)=\Delta TF_{\text{S}}(T_{0\text{K}},P_{0},\Delta
T,\Delta\xi); \label{NESCL_Entropy_Series}%
\end{equation}
we have not shown any dependence on $\xi_{\text{eq,K}}$ as it is a function of
$T_{0\text{K}},P_{0}$\ so it is no longer independent. The extensive function
takes the value $F_{S}(T_{0\text{K}},P_{0},0,0)=C_{P\text{K}}/T_{0\text{K}}$.
It must increase as $\Delta T$decreases during isothermal relaxation to ensure
that $S_{\text{NESCL}}$ continues to increase.

A similar Taylor expansion can be made for the volume $V_{\text{NESCL}}$. We
first recall that $V_{\text{f,NESCL}}$ and the communal entropy
$S_{\text{NESCL}}$ vanish simultaneously\footnote{In contrast, there is no
such relationship between the configurational entropy and the free volume. The
most simple way to appreciate is to recognize that the configurational entropy
in the Gibbs-Di Marzio theory can vanish even when the polymers cover the
entire lattice so that the free volume is identically zero.} and that
\begin{equation}
S_{\text{NESCL}}(T,P_{0},\xi)=\sigma(T,P_{0},\xi)V_{\text{f,NESCL}}%
(T,P_{0},\xi). \label{NESCL_Entropy_Volume_Relation}%
\end{equation}
Thus, $V_{\text{f,NESCL}}$ is also non-singular and will have a Taylor
expansion around the Kauzmann point. To determine the nature of the expansion,
we follow the above approach to determine the leading powers of $\Delta T$and
$\Delta\xi$. The volume expansion coefficient at K due to the free volume is
nonzero as we approach K from the high temperature side because of the
difference in the slopes at K in Fig. \ref{Fig_Glass_Transition_Time}. Thus,
the expansion must start with a term linear in $\Delta T$. The determination
of the leading power of $\Delta\xi$requires considering the behavior of
$\left(  \partial V/\partial\xi\right)  _{T,P}$. It follows from the theorem
of small increments \cite{Landau} that%
\[
\left(  \partial V/\partial\xi\right)  _{T,P}=\left(  \partial^{2}\widehat
{G}/\partial\xi\partial P\right)  =\left(  \partial A/\partial P\right)
_{\xi,P},
\]
where $\widehat{G}$ is the double Legendre transform $E-TS+PV$%
\footnote{$\widehat{G}$ should not be confused with the Gibbs free energy
$G(t)=E(t)-T_{0}S(t)+P_{0}V(t)$. This point is carefully discussed in Ref.
\cite{Gujrati-I}.} so that $V=\left(  \partial\widehat{G}/\partial P\right)
_{\xi,T}$. As the affinity $A$ always vanishes at K regardless of the
pressure, the derivative on the right side must vanish. This means that
$\left(  \partial V/\partial\xi\right)  _{T,P}$ must vanish at K, which leads
to a leading bilinear combination $\Delta T\Delta\xi$. Thus, the situation is
as before for $S_{\text{NESCL}}$ so that we can write a similar form for
$V_{\text{f,NESCL}}$%
\begin{equation}
V_{\text{f,NESCL}}(T,P_{0},\xi)=\Delta TF_{\text{V}}(T_{0\text{K}}%
,P_{0},\Delta T,\Delta\xi). \label{NESCL_Volume_Series}%
\end{equation}
\qquad

We can use this form of the free volume in the Doolittle equation
(\ref{Doolittle Eq}) to obtain%
\begin{equation}
\eta=\eta_{0}\exp(\gamma v_{\text{m}}/(T-T_{0\text{K}})f_{\text{V}})
\label{General eta-T relation}%
\end{equation}
in terms of the intensive function $f_{\text{V}}\equiv F_{\text{V}}/N$. It
should be noted that all quantities above are functions of $T,P_{0}$ and $\xi
$. Using the linear relationship from (\ref{NESCL_Entropy_Volume_Relation}) in
(\ref{Doolittle Eq}), we obtain%
\begin{equation}
\eta=\eta_{0}\exp(\gamma v_{\text{m}}/gs_{\text{NESCL}})
\label{general eta-entropy relation}%
\end{equation}
in terms of the per particle communal entropy $s_{\text{NESCL}}\equiv
S_{\text{NESCL}}/N$. Again, all quantities are functions of state variables.

\subsubsection{Free Volume Theory}

The free volume theory of Doolittle \cite{Doolittle,Cohen-Grest} attempts to
describe \textit{both} aspects with some respectable success even though
lately it has fallen out of favor. This is unfortunate as this theory captures
the essence of the GT, which in this theory occurs when the free volume
becomes sufficiently \emph{small }to impede the \emph{mobility} of the
molecules \cite{Doolittle}. The Doolittle equation correctly predicts the
abrupt increase in the viscosity for a large number of glass formers in a
narrow range over which $v_{\text{f}}$ becomes very small; see ref.
\cite{Cohen-Grest} for further details. The time-dependence of the free-volume
redistribution, determined by the energy barriers encountered during
redistribution, should provide a \emph{kinetic} view of the transition, and
must be properly accounted for. This approach is yet to be completed to
satisfaction. Nevertheless, assuming that the change in free volume is
proportional to the difference in the temperature $T_{0}-T_{0\text{V}}$ near
some temperature $T_{0\text{V}}$, it is found that $\eta(T_{0})$ diverges near
$T_{0V}$ according to the phenomenological Vogel-Tammann-Fulcher equation%
\begin{equation}
\ln\eta(T_{0})=A_{\text{VTF}}+B_{\text{VTF}}/(T_{0}-T_{0\text{V}}),
\label{eta1}%
\end{equation}
where $A_{\text{VTF}}$ and $B_{\text{VTF}}$ are system-dependent constants. A
comparison with (\ref{General eta-T relation}) shows the phenomenological
equation to be a special case of the general equation. The limited validity of
the original (constant parameters) Doolittle equation also makes the VTF
equation with limited validity. in addition, different concept of the free
volume will also yield different temperatures where it vanishes. This explains
the puzzling differences between $T_{0\text{V}}$ and the Kauzmann temperature
noted by several workers.

\subsubsection{Adam-Gibbs Theory}

The thermodynamic theory due to Adam and Gibbs \cite{Adam-Gibbs,Gujrati-Book}
attempts to provide a justification of the \emph{entropy crisis} in SCL. The
central idea is that the sluggishness observed in a system is a manifestation
of the smallness of the configurational entropy, i.e. the smallness of the
available configurations to the system. The configurational entropy in this
theory is identical to the configurational entropy $S^{\text{conf}}(T_{0})$
defined above in (\ref{CK-Partition}). This entropy should not be confused
with the communal entropy that we have been considering above. According to
this theory, the viscosity $\eta(T_{0})$ above the glass transition is
given\ as follows:%
\begin{equation}
\ln\eta(T_{0})=A_{\text{AG}}+B_{\text{AG}}/T_{0}s^{\text{conf}}, \label{eta2}%
\end{equation}
where $A_{\text{AG}}$ and $B_{\text{AG}}$ are system-dependent constants and
$s^{\text{conf}}=S^{\text{conf}}/N$. It is commonly believed that the
configurational entropy also vanishes at a positive temperature $T_{0\text{S}%
}$, which is not identical to $T_{0\text{K}}$, although they are found to be
close \cite{Cohen-Grest}. A similar argument as used above will also show
that
\[
S_{\text{NESCL}}^{\text{conf}}(T,P_{0},\xi)=\Delta T_{\text{S}}F_{\text{S}%
}^{\text{conf}}(T_{0\text{V}},P_{0},\Delta T_{\text{V}},\Delta\xi),
\]
where $\Delta T_{\text{S}}(t)\equiv T(t)-T_{0\text{S}}$. The derivation of
(\ref{eta2}) is based on the concept of cooperative domains of size $z$, which
gradually increases as the temperature is lowered and diverges at
$T_{0\text{S}}$. At this temperature, the entire system acts like a
cooperative domain. While this particular domain is disordered, its
configurational entropy must vanish in this theory. As the laboratory glass
transition temperature $T_{0\text{g}}$ occurs at about 50\ K above
$T_{0\text{V}}$, the value of $z$ at $T_{0\text{g}}$ is much smaller; it is
found to be of the order of $5-10$.

If we replace $s^{\text{conf}}$ in (\ref{eta2}) by the communal entropy, then
the above equation represents a special case of the general thermodynamic
equation given in (\ref{general eta-entropy relation}) except that the
viscosity diverges at different temperatures in the two approaches. As the two
theories are developed based on different approaches but with the same
conclusion, it appears that the suggestion of a rapid rise\ in the viscosity
due to a sudden drop in some form of entropy seems very enticing, since both
phenomena are ubiquitous in glassy states. Thus, we are driven to the
conclusion that we can treat the SCL glass transition within a thermodynamics
formalism in which some sort of entropy crisis is exploited. While the
vanishing of free volume is not tied to the vanishing of $s^{\text{conf}}$,
the vanishing of free volume occurs simultaneously with the vanishing of the
communal entropy. Thus, our approach exploiting the communal entropy ties the
divergence of the viscosity with the vanishing of either the free volume or
the communal entropy at the same temperature $T_{0\text{K}}$.

\subsection{Mode-Coupling Theory}

The mode-coupling theory \cite{Gotze} is an example of theories based on
kinetic ideas in which deals not with the glass transition but with the
transition at $T_{\text{MC}}.$ Thus, it is not directly relevant \ for our
review. This theory may be regarded as a theory based on first-principle
approach, which starts from the static structure factor. In this theory, the
ergodicity is lost completely, and structural arrest occurs at a temperature
$T_{\text{MC}}$, which lies well above the customary glass transition
temperature $T_{\text{G}}$. Consequently, the correlation time and the
viscosity diverge due to the \emph{caging effect}. The diverging viscosity can
be related to the vanishing free volume \cite{Cohen-Grest,Doolittle}, which
might suggest that \emph{the MC transition is the same as the glass
transition}. This does not seem to be the consensus at present. It has been
speculated that the divergence at $T_{\text{MC}}$ is due to the neglect of any
activated process in this theory. This, however, has been disputed by recent
studies. The mode-coupling theory is also not well-understood, especially
below the glass transition. More recently, it has been argued that this and
mean-field theories based on an underlying first-order transition may be
incapable of explaining dynamic heterogeneities.

\subsection{Random First Order Theory}

An alternative thermodynamic theory for the impending entropy crisis based on
spin-glass ideas has also been developed in which proximity to an underlying
first-order transition is used to explain the glass transition \cite{Wolynes}.
The theory is based on the one-step replica symmetry breaking in spin glasses
which also undergo a spin-glass transition similar in many respect to the
freezing transition in glasses. The one-step replica symmetry breaking is
identified in a long-range spin glass model so the theory is at a mean field
level. While some may consider this to be a weakness of the theory, it also
provides a new level of intuition about ordinary glasses.

\section{A Simple Model of a Non-equilibrium Temperature\label{Sect_Model}}

The possibility of a temperature disparity can be heuristically demonstrated
by considering a simple non-equilibrium\ laboratory problem. Consider a system
as a \textquotedblleft black box\textquotedblright\ consisting of two parts at
different temperatures $T_{1}$and $T_{2}>T_{1}$, but insulated from each other
so that they cannot come to equilibrium. The two parts are like slow and fast
motions in a glass, and the insulation allows us to treat them as independent,
having different temperatures. We assume that there are no irreversible
processes that go on within each part so that there is no irreversible heat
$d_{\text{i}}Q_{1}$ and $d_{\text{i}}Q_{2}$ generated within each part. We
wish to identify the effective temperature of the system. To do so, we imagine
that each part is added a certain \emph{infinitesimal} amount of heat from
outside, which we denote by $dQ_{1}$ and $dQ_{2}.$ The amount of heat $dQ$
added to the system is their sum. We assume the entropy changes to be $dS_{1}$
and $dS_{2}$. Then, we have
\[
dQ=dQ_{1}+dQ_{2},dS=dS_{1}+dS_{2}.
\]
Let us introduce a temperature $T$ by $dQ=TdS$. Using $dQ_{1}=T_{1}%
dS_{1},dQ_{2}=T_{2}dS_{2}$, we immediately find

\qquad\qquad\qquad%
\[
dQ(1/T-1/T_{2})=dQ_{1}(1/T_{1}-1/T_{2}).
\]
By introducing $x=dQ_{1}/dQ$, which is determined by the setup, we find that
$T$ is given by%
\begin{equation}
\frac{1}{T}=\frac{x}{T_{1}}+\frac{1-x}{T_{2}}. \label{T_eff}%
\end{equation}
As $x$ is between $0$ and $1$, it is clear that $T$ lies between $T_{1}$ and
$T_{2}$ depending on the value of $x$. Thus, we see from this heuristic model
calculation that the effective temperature of the system is not the same as
the temperature of either parts, a common property of a system not in equilibrium.

The above calculation is for fixed $T_{1}$ and $T_{2}$ since the infinitesimal
heats do not change the temperatures and there is no energy exchange between
the two parts due to insulation. It is the value of $x$ that uniquely
determines the temperature $T$ of the system. It depends on the way the two
heats are exchanged.

If the insulation between the parts is not perfect, there is going to be some
energy transfer between the two parts, which would result in maximizing the
entropy of the system. As a consequence, their temperatures will eventually
become the same. During this time, the temperature $T$ of the system will lie
between the changing temperatures of the two parts, and will itself be changing.

\section{The Fictive Temperature and the Tool-Narayanaswamy
Equation\label{Marker_Fictive}}

We will exhibit $t$ in this section.

\subsection{Partitioning the Degrees of freedom (dof)}

The relaxation of all thermodynamic properties in the temporal domain depends
on the of the state of the system. For example, at high enough temperatures,
the time variation of $T(t)$ as it relaxes towards $T_{0}$ can be described as
a single simple exponential with a characteristic time scale $\tau$, the
relaxation time. This happens when all the degrees of freedom (dof) come to
equilibrium simultaneously with the same relaxation time. At any time $t$
before equilibrium is reached, the system has a temperature $T(t)$; it also
has energy $E(t),V(t),\xi(t)$ etc.

At low temperatures, this is not true. There are slow and fast modes noted in
Sect. \ref{marker-Sec-Introduction}. The situation is similar to the simple
system considered in Sect. \ref{Sect_Model}, which we now imagine to be in a
medium at temperature $T_{0}$. Both parts will strive to come to equilibrium
with the medium but they may have widely separated relaxation times describing
the fast and slow modes in the system. Situation similar to this also occurs
in the attainment of thermal equilibrium between the nuclear spins and their
environment during nuclear relaxation studied by Purcell, where the
spin-lattice relaxation is extremely slow. The explanation of the slow and
fast dof in a wide class of substances lies in internal molecular motions
other than simple vibrations. The fast dof cool down and equilibrate very
fast, while the slow dof take much longer to transfer their energy and
equilibrate because of very weak coupling with the surrounding medium. We will
denote those dof that have equilibrated with the medium at time $t$ by a
subscript "e", and the remaining that are not equilibrated by "n."

Let $D$ denote the total number of the dof in the system, which is determined
by the number of particles $N$ in it; hence, it remains constant. The fast dof
equilibrate within the observation time $t_{\text{obs}},$\ with the slow dof
remaining out of equilibrium \cite{Goldstein}.\ Eventually, as $t\rightarrow
\tau_{\text{eq }}$, all dof come to equilibrium with the medium. Let
$D_{\text{e}}(t)$ and $D_{\text{n}}(t)$ denote its partition in equilibrated
and non-equilibrated dof, respectively:%
\[
D=D_{\text{e}}(t)+D_{\text{n}}(t);
\]
evidently, they are functions of time. Assume that at $t=0$ the system is
cooled instantaneously from an equilibrated supercooled liquid at
$T_{0}^{\prime}$ to a glass state at $T_{0}$. Immediately prior to cooling,
all dof are in equilibrium at $T_{0}^{\prime}$ and $D_{\text{e}}(t)=D$ for
$t\rightarrow0^{-}$. At $t=0$, all dof are out of\ equilibrium with the new
medium at $T_{0}$ so that $D_{\text{n}}(0)=D$. Eventually, $D_{\text{e}}(t)=D$
for all $t\geq\tau_{\text{eq}}$. It is clear that $D_{\text{e}}(t)$ does not
remain constant. Thus slow dof become part of $D_{\text{e}}(t)$ in time,.

The weak coupling between the two dof and of the slow dof with the medium
allows us to treat them as almost uncorrelated and \emph{quasi-independent},
which then immediately leads to the following partition of the $S,E,V$ and
$\xi$ into two contributions, one from each kind:
\begin{equation}
\mathbf{Z}(t)=\mathbf{Z}_{\text{e}}(t)+\mathbf{Z}_{\text{n}}(t).
\label{Partitions_S}%
\end{equation}
It should be noted that\ $S_{\text{e}}(t)$\ and $S_{\text{n}}(t)$\ stand for
$S_{\text{e}}(E_{\text{e}}(t),V_{\text{e}}(t),\xi_{\text{e}}(t))$ and
$S_{\text{n}}(E_{\text{n}}(t),V_{\text{n}}(t),\xi_{\text{n}}(t)).$

We have already remarked earlier that $V_{\text{i}}$ in
(\ref{Volume-Partition}) corresponds to $D_{\text{e}}$ during $\tau
_{\text{obs}}$. During this time, the glass is trapped within an inherent
structure IS$_{0}$. The slow dof correspond to the center of mass motion
within the cells and to visit other ISs. They are too slow to equilibrate at
the new temperature. A large body of simulations discussed in Ref.
\cite{Gujrati-I} has established a very clear pattern for the mean square
displacement of a particle as a function of time $t$, starting from a
ballistic regime to a plateau to a final diffusive regime; see also Zallen
\cite{Zallen}.

\subsection{Fictive State and Temperature}

Let us now introduce%
\begin{equation}
x(t)\equiv dE_{\text{e}}(t)/dE(t),\ 1-x(t)\equiv dE_{\text{n}}(t)/dE(t),\ ,
\label{x_definition}%
\end{equation}
at a given $t$, so that
\begin{align}
\partial S_{\text{e}}(t)/\partial E(t)  &  =x(t)\partial S_{\text{e}%
}(t)/\partial E_{\text{e}}(t),\ \label{x_T_derivatives}\\
\partial S_{\text{n}}(t)/\partial E(t)  &  =[1-x(t)]\partial S_{\text{n}%
}(t)/\partial E_{\text{n}}(t).
\end{align}
At $t=0$, $D_{\text{e}}=0,$ $E_{\text{e}}=0$ and $x=0.$ At $t\geq
\tau_{\text{eq}}$, $D_{\text{e}}=D,\ E_{\text{n}}=0$ and $x=1$. As time goes
on, more and more of the "n" dof equilibrate, thus increasing $D_{\text{e}%
}(t)$ and $x(t)$.\qquad

By definition, we have $\partial S_{\text{e}}(t)/\partial E_{\text{e}%
}(t)=1/T_{0}$, while the dof$_{\text{n}}$ will have a temperature different
from this. Assuming internal equilibrium, we can introduce a new temperature
$T_{\text{n}}(t)$ by%
\begin{equation}
\partial S_{\text{n}}(t)/\partial E_{\text{n}}(t)=1/T_{\text{n}}(t).
\label{Fictive_Temp}%
\end{equation}
The following identity%
\begin{equation}
\frac{1}{T(t)}=\frac{x(t)}{T_{0}}+\frac{1-x(t)}{T_{\text{n}}(t)}
\label{Narayanaswamy_Decomposition}%
\end{equation}
easily follows from considering $\partial S(t)/\partial E(t)$ and using
(\ref{Partitions_S}) and (\ref{x_T_derivatives}). This equation should be
compared with (\ref{T_eff}) obtained earlier using a heuristic model.
Initially, $x(0)=0$ so that $T(0)=$ $T_{\text{n}}(0)=T^{\prime}$, while
$T(t)\rightarrow T_{0}$ as $t\rightarrow\tau_{\text{eq}}$, as expected. This
division of the instantaneous temperature $T(t)$ into $T_{0\text{ }}$and
$T_{\text{n}}(t)$ is identical in form to that suggested by Narayanaswamy
\cite{Goldstein}, except that we have given thermodynamic definitions of
$x(t)$ in (\ref{x_definition}) and $T_{\text{n}}(t)$ in (\ref{Fictive_Temp}).

\subsubsection{Physical Significance}

Let us now understand the significance of the above analysis. The partition in
(\ref{Partitions_S}) along with the fraction $x(t)$ shows that the partition
satisfies a lever rule: the relaxing glass can be \emph{conceptually} (but not
physically) thought of as a "mixture" consisting of two different "components"
corresponding to dof$_{\text{e}}$ and dof$_{\text{n}}$: the former is at
temperature $T_{0}$ and has a weight $x(t)$; the latter with a complementary
weight $1-x(t)$ is at a temperature $T_{\text{n}}(t).$ Thinking of a system
conceptually as a "mixture" of two "components" is quite common in theoretical
physics. One common example is that of a superfluid, which can be thought of
as a "mixture" of a normal viscous "component" and a superfluid "component".
In reality, there exist two simultaneous motions, one of which is "normal" and
the other one is "superfluid". A similar division can also be carried out in a
superconductor: the total current is a sum of a "normal current" and a
"superconducting current".

\subsubsection{Nonequilibrium Aspects and A Fictive State}

However, because of the non-equilibrium nature of the system, there is an
important difference between a glass and a superfluid or a superconductor. The
e-component is in equilibrium (with the medium), but the n-component is only
in internal equilibrium. While the significance of the former as a ESCL
"component" at $T_{0},P_{0}$ (dof=$D_{\text{e}}$) is obvious, the significance
of the latter requires clarification. At $t=0$, $T_{\text{n}}(t)=T_{0}%
^{\prime}$ of ESCL (dof=$D$) from which the current glass is obtained so that
$S_{\text{n}}(t=0)=S_{\text{ESCL}}(T_{0}^{\prime})$. Also, $E_{\text{n}%
}(t=0)=E_{\text{ESCL}}(T_{0}^{\prime}),V_{\text{n}}(t=0)=V_{\text{ESCL}}%
(T_{0}^{\prime}),$ and $\xi_{\text{n}}(t=0)=\xi_{\text{ESCL}}(T_{0}^{\prime
}).$ At any later time $t>0$, $T_{\text{n}}(t)$ represents the temperature
associated with the energy $E_{\text{n}}(t)$ and volume $V_{\text{n}}(t)$ of
the non-equilibrated "component" of the glass and has a weight $1-x(t)$. This
"component," being in internal equilibrium, can be identified as a
\emph{fictive} ESCL [dof=$D_{\text{n}}(t)$]\ at temperature $T_{\text{n}%
}<T_{0}^{\prime}$ of energy $E_{\text{ESCL}}=E_{\text{n}}(t),V_{\text{ESCL}%
}=V_{\text{n}}(t),$ and $\xi_{\text{ESCL}}=\xi_{\text{n}}(t).$ In other words,
the relaxing glass at any time $t$ can be considered as consisting of two ESCL
"components," one at temperature $T_{0}$ [dof=$D_{\text{e}}(t)$] and the other
one [dof=$D_{\text{n}}(t)$] at temperature $T_{\text{n}}(t)$. The temperature
$T_{0\text{f}}\equiv T_{\text{n}}(t)$ \emph{uniquely} determines
$E_{\text{ESCL}}(T_{0\text{f}})\equiv E_{\text{n}}(t),V_{\text{ESCL}%
}(T_{0\text{f}})\equiv V_{\text{n}}(t),$ and $\xi_{\text{ESCL}}(T_{0\text{f}%
})\equiv\xi_{\text{n}}(t)$ of the corresponding \emph{fictive} ESCL
[dof=$D_{\text{n}}(t)$], which is in \emph{equilibrium} with a medium at
temperature $T_{0\text{f}}$. Consequently, $\xi_{\text{ESCL}}(T_{0\text{f}})$
is no longer an independent state variable for the fictive ESCL.

As the above fictive liquid at $T_{0\text{f}}\equiv T_{\text{n}}(t)$ contains
only (or mostly) the slow dof, it does not yet really represent a ESCL
associated with the system at $T_{0\text{f}}$, as the former lacks
dof$_{\text{e}}$, while the latter contains all dof. This does not pose any
problem as the missing dof$_{\text{e}}$ at $T_{0\text{f}}$ are in equilibrium
not only with the dof$_{\text{n}}$ at $T_{0\text{f}}$, but also with a medium
at $T_{0\text{f}}$. Thus, one can consider "adding" these missing
dof$_{\text{e}}$ (dof=$D_{\text{e}}$) to the fictive liquid, which now
represents the equilibrated ESCL (dof=$D$) at $T_{0\text{f}}$. This ESCL is
not the same as the glass with its fictive $T_{\text{n}}(t)$, as the latter
has its dof$_{\text{e}}$ at $T_{0}$ while the ESCL has all of its dof at
$T_{\text{n}}(t)$. However, all of their thermodynamic properties associated
with dof$_{\text{n}}$ must be the same, as their entropy function is the same
for both liquids. Similarly, the ESCL "component" at $T_{0},P_{0}$
(dof=$D_{\text{e}}$) should also be "supplemented" by the missing
dof$_{\text{n}}$ to give rise to the equilibrated ESCL at $T_{0},P_{0}$
(dof=$D$).

\subsubsection{Fictive Temperature}

We are now in a position to decide which of the temperatures $T(t)$ and
$T_{\text{n}}(t)$ qualifies as the \emph{fictive temperature}. We will
identify this temperature to characterize only the non-equilibrated dof (with
respect to the medium, but having internal equilibrium among themselves) in
the system, though other definitions are also possible. As $T(t)$ contains
information about both kinds of dof, it is not the appropriate temperature to
be identified as the fictive temperature. The temperature $T_{\text{n}}(t)$,
on the other hand, depends only on non-equilibrated dof$_{\text{n}}$,\ and
should be identified as the fictive temperature of the relaxing glass at time
$t$. This temperature is not the instantaneous temperature of the glass at
this time, but represents the equilibrium temperature of the corresponding
ESCL at $T_{0\text{f}}\equiv T_{\text{n}}(t)$.

\subsection{Tool-Narayanaswamy Phenomenology:\ Single Slow Relaxation}

The viscosity keeps changing with time during relaxation but it remains the
property of the system. Thus, it must depend not on $T_{0}$, but on $T(t)$,
the instantaneous temperature that characterizes the instantaneous state of
the glass. Using an Arrhenius form for the viscosity, we have
\begin{equation}
\eta(t)=\eta_{0}\exp\left[  B\left(  \frac{x(t)}{T_{0}}+\frac{1-x(t)}%
{T_{0\text{f}}(t)}\right)  \right]  , \label{Narayanaswamy_Relaxation}%
\end{equation}
the form conventionally identified as the phenomenological Tool-Narayanaswamy
equation \cite{Goldstein}. Here, $\eta_{0}$ and $B$ are some parameters of the
system. From the discussion in Sec.
\ref{marker_Sec_Thermodynamic Considerations}, they must in general depend on
$T_{0},P_{0},T_{0\text{f}}(t)$ and the affinity $A(t)$. \ \ \ \ \ \ \ \ \ \ \ \ \ \ \ \ \ \ \ \ \ \ \ \ \ \ \ \ \ \ 

It should be noted that our definition of the fictive temperature
$T_{\text{n}}(t)$ makes it somewhat different from the conventional definition
used in the literature \cite{Goldstein}, which takes different values for
different quantities such as the enthalpy and the volume. While we do not
discuss it here, we have discussed it elsewhere \cite{Gujrati-I} that the
above definition of the fictive temperature and $x(t)$ is the same even if we
use the partition of the volume instead of the energy. Therefore, we need to
follow the consequences of this difference in their definition. This will
require a particular model of the dynamics in the system. A common acceptable
form is the Kohlrausch form in (\ref{Kohlrausch-Form}). The exponential itself
may be taken to be a function of time and temperature to account for
deviations seen at short times $\left[  x(t)\simeq0\right]  $ and long times
$\left[  x(t)\simeq1\right]  $. Usually, $\beta$ increases monotonically with
the temperature. Thus, it will also change during relaxation as $T(t)$
changes. Let us assume for the moment that our $T_{\text{n}}(t)$ is not very
different from the customary fictive temperature. Conventionally, the
viscosity is fitted by taking $x$ as a constant close to $0.5$, but allowing
three other adjustable parameters ($\eta_{0},B,$ and $\beta$) to obtain the
best fit \cite[see the contribution by Moynihan et al]{Goldstein}; all $4$
parameters will generally have some time-dependence, but their time-dependence
is neglected in finding the best fit. Indeed, even the values of the fictive
temperature have appreciable uncertainties depending on the procedure to find
it. Therefore, such fits do not rule out a slowly varying $x(t)$.
Time-dependence of $x(t)$ has been recognized for quite some time in the
literature; see Ref. \cite{Gujrati-I} for more details.

There cannot be any doubt that a constant $x$ in
(\ref{Narayanaswamy_Relaxation}), that is $x$ being independent of the aging
conditions, is an approximation when used to describe experiments. But this is
most certainly not correct as no nonequilibrium state, in which $D_{\text{e}%
}(t)$ and $D_{\text{n}}(t)$ have different temperatures,\ can be identified
with an equilibrium state with all $D$ dof at temperature $T_{0\text{f}}(t)$.
Recall that there is a unique relationship between $E_{\text{ESCL}%
}[T_{0\text{f}}]=E(t)$ and the temperature $T_{0\text{f}}$.\ However, there
can be a variety of glasses with different energies but all having the same
fictive temperature $T_{0\text{f}}$. Thus, the original idea of Tool cannot be
correct. What our approach shows is that an ageing glass has two distinct dof
and only the non-equilibrated dof should be identified with the equilibrated
liquid at $T_{\text{n}}(t)$. This picture now no longer supports
ageing-independent $x(t)$. This is where our new understanding differs from
the original idea of Tool. This also makes data-fitting a challenge. This is
the price to be paid for changing $x$ from an empirical parameter to a
thermodynamic quantity. However, the benefit of our approach is that the
fictive temperature is the same whether we consider the energy or the volume.
It would be interesting to see what kind of time- and temperature-dependence
$x(t)$ will exhibit with our definition $T_{\text{n}}(t)$ of the fictive
temperature. This will require introducing a particular dynamics, which is not
our aim in this paper.

The extension to more than one slow relaxation has been considered by us in
Ref. \cite{Gujrati-I} and will not be pursued here.

\section{Discussion and Conclusions}

After giving a brief review of some of the important issues in vitrification
and its phenomenology, we follow it up with a brief introduction to a recently
developed non-equilibrium thermodynamics of a system in internal equilibrium
and apply it to supercooled liquids and glasses. The concept of internal
equilibrium requires the system to be homogeneous and its instantaneous
entropy to be maximum for the state variables at that instant. The state
variables include some internal variables that cannot be controlled by the
observer. An inhomogeneous system also requires internal variables. A simple
example of an internal variable for an inhomogeneous system is given, which is
again considered to introduce the instantaneous temperature later in Sec.
\ref{Sect_Model}. This model is central to justify the Tool-Narayanaswamy
equation later.

The conventional approaches to study viscosity as a function of temperature
either uses $\Delta S_{\text{ex}}$\ or $V_{\text{f}}$. The Adam-Gibbs theory
is based on $S^{\text{conf}}$ instead of $\Delta S_{\text{ex}}$.\ (However,
the two are not the same, at least for polymers. While the former contains the
conformational entropy, it is absent in $\Delta S_{\text{ex}}$.) By replacing
$S^{\text{conf}}$ by $\Delta S_{\text{ex}}$, one can determine the temperature
$T_{0\text{S}}$ where the viscosity diverges in the Adam-Gibbs theory. In the
free volume theory, the viscosity diverges at $T_{0\text{V}}$. The two
temperatures are usually different as there is no relationship between the
vanishing of $V_{\text{f}}$ and $\Delta S_{\text{ex}}$. This is puzzling as
the state of the system is its thermodynamic property and is independent of
the theory used to describe the system. Moreover, the state with a diverging
viscosity must be a unique state in that once the viscosity has diverged, it
cannot change in time. In general, for a system in internal equilibrium, the
viscosity must be a function of the state variables:$\eta=\eta
(T(t),P(t),A(t)).$ For a state with diverging viscosity, there cannot be any
variations in the fields. In other words, we expect a unique temperature where
$\eta$ diverges so that the above two temperatures must be the same. This
mismatch is remedied by our approach in which we take the ideal glass state to
be a unique state in which $S^{\text{comm}}$ and $V_{\text{f}}$ vanish
together so that the above two temperatures are not different from the
Kauzmann temperature $T_{0\text{K}}$. Thus, the different looking (free volume
and Adam-Gibbs) theories become identical as we have shown. This unification
comes from the uniqueness of the ideal glass. Whether we ever get to the state
is not relevant for the mathematical expansion around the Kauzmann point.

The nonequilibrium nature of SCL and GS requires that we make a distinction
between the instantaneous fields and those of the medium. If this is not done,
as is usually the case in most nonequilibrium approaches in which internal
variables are introduced, then the Gibbs free energy does not change, while it
must decrease during relaxation for a system out of equilibrium. Indeed,
$T(t)-T_{0},P(t)-P_{0}$, etc. act as thermodynamic forces that drive the
system towards equilibrium. These forces have some important consequences for
how fields like $T(t)$, etc. and thermodynamic quantities like the volume,
entropy, enthalpy, etc. change in time.

The actual form of the dynamics in time was not considered here as our
interest was to understand how thermodynamic quantities change with fields.
However, the nature of the dynamics was incorporated in an indirect way be
realizing that the dynamics in SCL and GS should be divided into fast and slow
dynamics. Based on this observation, it was necessary to divide $V$ and $S$
into two parts, depending on the fast modes ($V_{\text{i}}$ and $S^{\text{int}%
}$) and slow modes ($V_{\text{f}}$ and $S^{\text{comm}}$). General
considerations show that these quantities are linearly related. As a
consequence, $V_{\text{f}}$ and $S^{\text{comm}}$ vanish simultaneously in IG.
As IG emerges out of ESCL, it is an equilibrium state in $\mathcal{D}$ so that
its fields are those of the medium. When the entropy of some NESCL is
extrapolated to energies below $E_{\text{G}}$, as was discussed in reference
to Fig. \ref{Fig.NE-E_SCL._entropy}, we argued that the extrapolated state of
zero communal entropy must be IG; the local oscillatory motion in the cages
are governed by equilibrium thermodynamics.

The linear relation $S^{\text{comm}}=V_{\text{f}}\sigma$ associated with the
free volume $V_{\text{f}}$\ is different from alternate choice $S_{\text{f}%
}=-\ln V_{\text{f}}$. We do not consider the latter choice as it gives
negative communal entropy, whereas we have required it to non-negative. \ \ 

We have clarified the concept of the fictive temperature $T_{\text{n}}(t)$
widely used in the study of glasses by identifying it as a thermodynamic
quantity; see (\ref{Fictive_Temp}). Our analysis shows that the fictive
temperature has the same value even if we change the relaxing quantity from
the energy to the volume. This temperature is not identical to but is related
to the instantaneous temperature $T(t)$ in a glass; see
(\ref{Narayanaswamy_Decomposition}). We use this relationship to establish the
Tool-Narayanaswamy equation (\ref{Narayanaswamy_Relaxation}) for the
relaxation time on a solid theoretical ground. As we have been able to offer a
thermodynamic interpretation of $x(t)$, it no longer is merely a parameter
following the original idea of Tool. Therefore, it is not surprising that the
time-dependence of $x(t)$ requires reinterpreting Tool's original idea of the
fictive temperature. It should not be interpreted as the fictive temperature
at which the glass is in equilibrium; rather, it is only the non-equilibrated
dof of the glass that is compared with the equilibrium liquid at the fictive
temperature $T_{\text{n}}(t)$.

The limitation of the paper should be mentioned. We have not discussed recent
work dealing with the heterogeneity in space and time to discuss glasses. The
reason for this is that the nonequilibrium thermodynamics that we are using
requires the additivity of the entropy for different parts. This requires the
parts to be macroscopically large so that surface effects can be neglected.
Thus, the approach is not applicable to a few particles for which we need
small size nonequilibrium thermodynamics, a field which is in infancy at present.


\begin{thebibliography}{99}                                                                                               %


\bibitem {Zallen}R. Zallen, \textit{The Physics of Amorphous Solids}, John
Wiley, New York\ (1983).

\bibitem {Goldstein}\textit{The Glass Transition and the Nature of the Glassy
State}, edited by M. Goldstein and R. Simha, N.Y. Academy of Sciences, New
York (1976).

\bibitem {Fischer-Hertz}K. Fischer and J. Hertz, \textit{Spin Glasses},
Cambridge University Press, Cambridge, U.K.; Reprint edition (1993).

\bibitem {Kauzmann}W. Kauzmann, Chem. Rev., \textbf{43}, 219-256 (1948).

\bibitem {Landau}L.D.\ Landau, E.M. Lifshitz, \textit{Statistical Physics},
Part 1, Third Edition, Pergamon Press, Oxford (1986).

\bibitem {Gujrati-Book}P.D. Gujrati in \textit{Modeling and Simulation in
Polymers}, edited by P.D. Gujrati and A.I. Leonov, Wiley-VCH, , Weinheim (2010).

\bibitem {Gujrati-Entropy}P.D. Gujrati, arXiv:1304.3768; P.D. Gujrati, Entropy
\textbf{17}, 710 (2015).

\bibitem {Ngai}K.L. Ngai in \textit{Soft Matter under Exogenic Impacts}, S.J.
Rzoska and V.A. Mazur, ed. p. 91, Springer (2007).

\bibitem {Matsuoka}S. Matsuoka, Poly. Eng. Sci. \textbf{21}, 907 (1981).

\bibitem {Doolittle}A.K. Doolittle, J. Appl. Phys. \textbf{22}, 1471 (1951).

\bibitem {Cohen-Grest}G.S. Grest and M.H. Cohen; Adv. Chem. Phys. \textbf{48},
455 (1981).

\bibitem {Nemilov}S.V. Nemilov,\ \textit{Thermodynamic and Kinetic Aspects of
the Vitreous State}, CRC, Boca Raton (1995).

\bibitem {Gujrati-I}P.D. Gujrati, Phys. Rev. E \textbf{81}, 051130 (2010);
P.D. Gujrati, arXiv:0910.0026.

\bibitem {Kovacs}J.J. Aklonis and A.J. Kovacs in \textit{Contemporary Topics
in Polymer Science}, M. Shen, ed. Vol. 3, p. 267 (1979).

\bibitem {Debenedetti}P.G. Debenedetti, \textit{Metastable Liquids, Concepts
and Principles}; Priceton University Press: Princeton, NY, USA (1996).

\bibitem {Debenedetti1}P.G. Debenedetti and F.H. Stillinger, Nature
\textbf{410}, 259 (2001).

\bibitem {Adam-Gibbs}G. Adams and J.H. Gibbs, J. Chem. Phys. \textbf{43}, 139 (1965).

\bibitem {Gotze}W. Gotze and L. Sjogren, Rep. Prog. Phys. \textbf{55}, 241(1992).

\bibitem {Wolynes}V. Lubchenko and P.G. Wolynes, Annu. Rev. Phys. Chem.
\textbf{58}:235 (2007).
\end{thebibliography}
\end{document}